\documentclass[aps,amssymb,amsmath,prd,twocolumn,
showpacs,preprintnumbers,superscriptaddress,nofootinbib,floatfix]{revtex4-1}

\usepackage[10pt]{type1ec}  
\usepackage[T1]{fontenc}
\usepackage{CJKutf8}
\usepackage[overlap, CJK]{ruby}
\usepackage{CJKulem}
\newenvironment{SChinese}{%
  \CJKfamily{gbsn}%
  \CJKtilde
  \CJKnospace}{}

\usepackage{graphicx}
\usepackage{bm}
\usepackage{mathrsfs}
\usepackage{latexsym}
\usepackage{float}
\usepackage{color}
\usepackage{dcolumn}
\usepackage[colorlinks=true,citecolor=blue,urlcolor=blue]{hyperref}
\usepackage[usenames,dvipsnames]{xcolor}
\usepackage[caption=false]{subfig}
\usepackage{slashed}

\newcommand\Eqn[1]{Eq.~(\ref{#1})}  

\newcommand{\p}{\partial}
\newcommand{\nn}{\nonumber}
\newcommand{\be}{\begin{equation}}
\newcommand{\ee}{\end{equation}}
\newcommand{\beq}{\begin{eqnarray}}
\newcommand{\eeq}{\end{eqnarray}}

\newcommand{\csqeq}{c_{\rm eq}^2}
\newcommand{\csqad}{c_{\rm ad}^2}
\newcommand{\ceq}{c_{\rm eq}}
\newcommand{\cad}{c_{\rm ad}}
\newcommand{\nb}{n_{\rm B}}
\newcommand{\nsat}{n_{\rm sat}}
\newcommand{\Msolar}{{\rm M}_{\odot}}
\newcommand{\Mmax}{\ensuremath{M_{\rm max}}}
\newcommand{\Rtyp}{R_{1.4}}

\newcommand{\Rmax}{\ensuremath{R_{\Mmax}}}
\newcommand{\Rskin}{R_\mathrm{skin}^{^{208}\mathrm{Pb}}}

\newcommand{\ep}{\varepsilon}

\newcommand{\bv}{Brunt--V\"ais\"al\"a}
\newcommand{\gm}{$g$-mode}
\newcommand{\gms}{$g$-modes}

\begin{document}
\preprint{N3AS-21-012, INT-PUB-21-021}
\title{$g$-modes of neutron stars with hadron-to-quark crossover transitions}

\author{Constantinos Constantinou}
\email{cconstantinou@ectstar.eu}
\affiliation{INFN-TIFPA, Trento Institute of Fundamental Physics and Applications, Povo, 38123 TN, Italy}
\affiliation{European Centre for Theoretical Studies in Nuclear Physics and Related Areas, Villazzano, 38123 TN, Italy}

\author{Sophia Han 
(\begin{CJK}{UTF8}{}\begin{SChinese}韩 君\end{SChinese}\end{CJK})
}
\email{sjhan@uw.edu}
\affiliation{Institute for Nuclear Theory, University of Washington, Seattle, WA~98195, USA}
\affiliation{Department of Physics, University of California, Berkeley, CA~94720, USA}

\author{Prashanth Jaikumar}
\email{prashanth.jaikumar@csulb.edu}
\affiliation{Department of Physics and Astronomy, California State University Long Beach, Long Beach, CA~90840, USA}

\author{Madappa~Prakash}
\email{prakash@ohio.edu}
\affiliation{Department of Physics and Astronomy, Ohio University,
Athens, OH~45701, USA}

\date{November 20, 2021}

\begin{abstract}

We perform the first study of the principal core \gm~oscillation in hybrid stars containing quark matter, utilizing a crossover model for the hadron-to-quark transition inspired by lattice QCD. The ensuing results are compared with our recent findings of \gm~frequencies in hybrid stars with a first-order phase transition using Gibbs constructions. 
We find that models using Gibbs construction yield \gm~amplitudes and the associated gravitational energy radiated that dominate over those of the chosen crossover model owing to the distinct behaviors of the equilibrium and adiabatic sound speeds in the various models. 
Based on our results, we conclude that were \gms~to be detected in upgraded LIGO and Virgo detectors it would indicate a first-order phase transition akin to a Gibbs construction.

\end{abstract}

\maketitle

\section{Introduction}
\label{sec:intro}

Dense matter inside neutron stars (NSs) could contain unbound quarks that retain some vestige of the forces described by the fundamental theory of the strong interaction, Quantum Chromodynamics (QCD).  
Lacking exact methods for its solution, numerous models have been constructed to investigate the hadron-to-quark transition in the region of high baryon density and zero temperature. On the other hand, the phase diagram of QCD at low density (zero or small baryon chemical potential) and high temperature is amenable to precision numerical studies which clearly point to a crossover with no clear phase boundary between the hadron resonance gas and the quark-gluon plasma~\cite{Borsanyi:2010cj}. 

Recently, Kapusta and Welle~\cite{Kapusta:2021ney} (KW hereafter) 
have proposed a crossover model for the hadron-to-quark transition in NSs to mimic the crossover feature of finite temperature lattice studies. The key trait of this model is an analytic mixing or switching function that accounts for the partial pressure of each  component as a function of a single parameter - the baryon chemical potential. They found that NSs as massive as $\sim2.2\,\Msolar$ could be supported by their crossover equation of state (EOS). 
As hadrons/nucleons and quarks both appear explicitly as separate degrees of freedom in the KW description, it is straightforward to keep track of their individual contributions to the total pressure. KW report that within their model between $1-10\%$ of the total pressure could be contributed by quark matter in the core.  
In treatments in which hadron and quark interactions are intermingled, the individual contributions from hadrons and quarks to the total pressure may not be possible to disentangle. 
Given that EOSs with first-order phase transitions treated using the Maxwell construction face some challenges in obtaining the stable $\sim 2\,\Msolar$ NSs that have been observed, it is worthwhile to investigate alternatives such as the crossover model for hybrid stars. 

Our objective in this paper is two-fold: first, we extend the pure neutron matter (PNM) model of KW~\cite{Kapusta:2021ney} 
to include $\beta$-equilibrium, as well as a crust for the hybrid star. We also include vector interactions among quarks. These modifications enable a direct comparison to other approaches in the literature~\cite{Han:2019bub}. Second, we investigate \gm~oscillations, a potentially observable signature of the hadron-to-quark transition. 
A \gm~is a specific fluid oscillation where a parcel of fluid is displaced against the background of a stratified environment inside a neutron star. While pressure equilibrium is rapidly restored via sound waves, chemical equilibrium can take longer, causing buoyancy forces to oppose the displacement. Since cold NSs are not convective, the opposing force sets up stable oscillations, with a typical frequency, called the (local) \bv~frequency. The kind of core \gms~we study here were introduced in~\cite{RG92,RG94,Lai:1993di,Lai:1998yc} and in a recent work, which we shall refer to as Paper I~\cite{Jaikumar:2021jbw}, we showed that the \gm~frequency rises steeply with the onset of quarks due to a rapid change in the equilibrium sound speed (see also Ref.~\cite{Wei:2018tts}). Since \gm~oscillations couple to tidal forces, they may be excited during the merger of two NSs and provide information on the interior composition, specifically quarks here. 
In Paper I, we chose a Gibbs construction for the mixed phase, which yields NS properties that are more compatible with astrophysical constraints than a Maxwell construction.\footnote{See Table VII in Ref.~\cite{Han:2019bub} for a comprehensive comparison of Gibbs, Maxwell and certain crossover models.} In this work, we present the systematics of the \gm~frequency for hybrid stars in a crossover scenario, adopting the generalized KW model as a representative of this class. 
It is worth noting that Maxwell-constructed first-order phase transitions cannot generate \gms~in the transition region, because the equilibrium and adiabatic sound speeds both become zero (due to frozen pressure and composition over the phase coexistence region) there. 

In the first phase of this work, we will restrict ourselves to zero temperature without the effects of superfluidity in both nucleons and quarks. 
To the best of our knowledge, this is the first study of \gm~for hybrid stars in a crossover model. References to earlier work in which hadronic EOSs with and without superfluidity were used to investigate \gm~frequencies can be found in Paper I (see Refs. [21]-[35] therein). 

Models akin to the crossover model of KW, but with some differences, have been considered earlier in the literature. Examples include the smooth crossover model of Ref.~\cite{Dexheimer:2014pea}, interpolated EOSs considered in Refs.~\cite{Baym:2017whm,Masuda:2012ed,Fukushima:2015bda,Kojo:2014rca}, and quarkyonic models of Refs.~\cite{McLerran:2018hbz,Zhao:2020dvu,Jeong:2019lhv,Sen:2020qcd}, et cetera. 
In the chiral model of Refs.~\cite{Papazoglou:1998vr,Dexheimer:2008ax,Dexheimer:2009hi}, a scalar field $\Phi$, acting as an order parameter, is responsible for the deconfinement phase transition which can be first-order or crossover depending upon temperature and baryon chemical potential. Depending on the specific EOS models used in the hadronic and quark phases, chemical potential and pressure equilibrium between the two phases - of either the Maxwell or the Gibbs sort - may not be realized~\cite{Baym:2017whm}. In such cases, several interpolation procedures have been adopted to connect the two phases on the premise that at high supra-nuclear densities, a purely hadronic phase is untenable. 

In quarkyonic models with a momentum shell structure~\cite{McLerran:2018hbz,Zhao:2020dvu}, 
quarks emerge at relatively low (but still 
supra-saturation) densities but remain bound by strong interactions below the Fermi surface: while hadrons and quarks are separated in momentum space, they coexist in configuration space. 
A key parameter that enters the calculation of the \gm, is
the squared adiabatic speed of sound $\csqad =(\partial P/\partial \ep)|_{y_i,\beta}$, where $P$ and $\ep$ are the total pressure and energy density, respectively, and $y_{i,\beta}$ refer to the partial fractions of each component in beta-equilibrium.\footnote{$\csqad$ is different from the squared equilibrium speed of sound $\csqeq = dP/d\ep$  commonly defined as $c_s^2$ in the literature for static EOS models.} 
We find that in 
the quarkyonic ``shell'' models of Refs.~\cite{McLerran:2018hbz,Zhao:2020dvu}, $\csqad$ becomes discontinuous 
with respect to density at the shell boundary. 
To address \gms~in this specific category of quarkyonic shell models, and to 
extend the study of NS oscillations to span the various ways in which quarks can affect the dense matter EOS and the properties of NSs, such discontinuities must be smoothed.
As such a task is outside the scope of the present work, we do not consider these quarkyonic models in this paper.

The organization of this paper is as follows. In Sec.~\ref{sec:EOS}, we describe the EOSs for pure hadronic matter, pure quark matter and leptons used in the construction of the 
KW crossover model of Ref.~\cite{Kapusta:2021ney} 
by extending it to include $\beta$-equilibrium and interactions between quarks. The rationale for our parameter choices and the basic features of this model are also highlighted here for orientation. 
Sec.~\ref{sec:multi_compt} 
reviews 
the thermodynamics of a multicomponent system as pertinent to the KW description of crossover matter generalized here to describe neutron-star matter (NSM). 
In Sec.~\ref{sec:KWG}, the KW model formulation of crossover matter is described followed by the procedure to render the unconstrained system\footnote{The unconstrained system here refers to matter in which baryon number conservation, charge neutrality and $\beta$-equilibrium are not imposed.} in $\beta$-equilibrium. This section also contains   
a comparison of the extended KW model with those of selected quarkyonic shell models. 
In Sec.~\ref{sec:soundspeed}, the calculation of the equilibrium and adiabatic sound speeds in the crossover and Gibbs approaches are outlined.  
In Sec.~\ref{sec:Results}, we present results for the chosen EOSs and their associated NS structural properties as well as the two speeds of sound in the crossover model, and discuss emergent differences from other models that include a phase transition. This section also contains results for the sound speed difference, the \bv~frequency, and the \gm~frequency in hybrid stars along with their interpretation. Our conclusions and outlook are in Sec.~\ref{sec:Concs}.

\section{Equation of State} 
\label{sec:EOS}

In the KW description of the crossover transition, EOSs in the pure hadronic and quark phases are combined using a mixing or switch function that depends on the baryon chemical potential (to be described below). 
We therefore begin by discussing EOS models that we use in each of these two sectors.  
As we extend the KW model to include leptons, the EOS in the leptonic sector is also provided. These EOSs are first set forth without reference to baryon number conservation, charge neutrality, as well as chemical equilibrium (i.e., unconstrained), which are imposed at the appropriate junctures. 

\subsection{Pure hadronic matter}

To describe nucleons, we use the Zhao-Lattimer (ZL) \cite{Zhao:2020dvu} parametrization of the EOS of neutron-star matter. With reasonable adjustments of its parameters, this EOS can be made consistent with laboratory data at nuclear saturation density $\nsat\simeq 0.16~{\rm fm}^{-3}$ as well as recent chiral effective field theory calculations 
of Ref.~\cite{Drischler:2020hwi,Drischler:2020fvz} in which error quantifications up to $\sim2.0\,\nsat$ were made. 
The high-density behavior can be controlled by varying the slope of the symmetry energy parameter, $L$, at $\nsat$ within the range established from analyses of nuclear and observational data, 
see, e.g., Ref.~\cite{Lattimer:2012xj}, and a power-law index. Consistency with astrophysical data on known masses and radii of NSs is also attainable with this EOS. 

To begin with, no constraints are placed on the multi-particle system and therefore the independent variables are the baryon density $\nb$ and the individual particle fractions $y_n$, $y_p$. The total energy density of nucleons with a common mass $m_H=939.5$ MeV is described by the ZL functional 
\beq
\ep_H &=& \frac{1}{8\pi^2\hbar^3}\sum_{h=n,p}\left\{k_{Fh}(k_{Fh}^2 + m_H^2)^{1/2}(2k_{Fh}^2 + m_H^2)\right.  \nn \\
 &-& \left. m_H^4\ln\left[\frac{k_{Fh}+(k_{Fh}^2 + m_H^2)^{1/2}}{m_H}\right]\right\} \nn \\
           &+& 4 \nb^2 y_n y_p \left\{\frac{a_0}{\nsat} 
            +\frac{b_0}{\nsat^{\gamma}} [\nb(y_n + y_p)]^{\gamma - 1}\right\}  \nn \\
            &+& \nb^2 (y_n - y_p)^2\left\{\frac{a_1}{\nsat} 
            + \frac{b_1}{\nsat^{\gamma_1}}[\nb(y_n + y_p)]^{\gamma_1-1}\right\}  \nn \\
\eeq
The Fermi momentum of nucleon species $h$\footnote{While the symbols ``$H$'' and ``$h$'' connote hadrons, they actually refer to nucleons in the context of this paper.} is given by $k_{Fh} = (3\pi^2\hbar^3\nb y_h)^{1/3}$. The constants $a_0, b_0$ and $\gamma$ refer to isospin symmetric matter, whereas $a_1, b_1$ and $\gamma_1$ to that of isospin asymmetric matter. The chosen values of these constants are listed in a later section.

The corresponding chemical potentials are
\beq
\mu_n &=& (k_{Fn}^2 + m_H^2)^{1/2}  \nn \\
  &+& 4 \nb y_p \left\{\frac{a_0}{\nsat}+ \frac{b_0}{\nsat^{\gamma}}[\nb(y_n + y_p)]^{\gamma-1}\right\} 
  \nn \\
  &+& 4 \nb^2 y_p y_n \frac{b_0}{\nsat^{\gamma}}(\gamma-1)[\nb(y_n + y_p)]^{\gamma-2} 
  \nn \\
 &+& 2 \nb (y_n-y_p)\left\{\frac{a_1}{\nsat} + \frac{b_1}{\nsat^{\gamma_1}}[\nb(y_n+y_p)]^{\gamma_1-1}\right\}  \nn  \\
   &+& \nb^2 (y_n-y_p)^2 \frac{b_1}{\nsat^{\gamma_1}}(\gamma_1-1)[\nb(y_n + y_p)]^{\gamma_1-2} \, ,   \\
\mu_p &=& (k_{Fp}^2 + m_H^2)^{1/2} \nn \\
  &+& 4 \nb y_n \left\{\frac{a_0}{\nsat}+ \frac{b_0}{\nsat^{\gamma}}[\nb(y_n + y_p)]^{\gamma-1}\right\} 
  \nn \\
  &+& 4 \nb^2 y_p y_n \frac{b_0}{\nsat^{\gamma}}(\gamma-1)[\nb(y_n + y_p)]^{\gamma-2} 
  \nn \\
 &-& 2 \nb (y_n-y_p)\left\{\frac{a_1}{\nsat} + \frac{b_1}{\nsat^{\gamma_1}}[\nb(y_n+y_p)]^{\gamma_1-1}\right\}  \nn   \\
   &+& \nb^2 (y_n-y_p)^2 \frac{b_1}{\nsat^{\gamma_1}}(\gamma_1-1)[\nb(y_n + y_p)]^{\gamma_1-2} \,.
\eeq
Note the opposite $y_h$ and signs in the second and fourth terms of $\mu_n$ and $\mu_p$, respectively. 

The pressure is obtained from the thermodynamic identity 
\be
P_H = \nb\sum_{h=n,p}\mu_h y_h - \ep_H
\ee 
and the equilibrium speed of sound from
\be
\left(\frac{
\ceq}{c}\right)^2 =   \frac {d P_H}{d  \ep_H} \,.
\ee
The adiabatic speed of sound is obtained by taking partial derivatives of the pressure and the total energy density with respect to baryon density while keeping all particle fractions fixed
\be
\left(\frac{
\cad}{c}\right)^2 = \left.\frac{\p P_H}{\p \nb}\right|_{y_h}
                           \left(\left.\frac{\p \ep_H}{\p \nb}\right|_{y_h}\right)^{-1}~.
\ee
This is made particularly convenient by the choice of starting with a completely unconstrained system. 

\subsection{Pure quark matter}

For the calculation of the quark EOS, we use the vMIT bag model~\cite{Gomes:2018eiv, Klahn:2015mfa}. The Lagrangian density of this model is
\be
\mathcal{L}=\sum_{q=u,d,s}\left[\bar{\psi}_{q}\left(i \slashed{\p} -m_{q}-B\right) \psi_{q}+\mathcal{L}_{\mathrm{vec}}\right] \Theta \,,
\ee
where $\mathcal{L}_{\mathrm{vec}}$ describes repulsive interactions between quarks of mass $m_{q}$ confined within a bag as denoted by the $\Theta$ function:
\be
\mathcal{L}_{\text {vec }}=-G_{v} \sum_{q} \bar{\psi} \gamma_{\mu} V^{\mu} \psi+\left(m_{V}^{2} / 2\right) V_{\mu} V^{\mu} \,.
\ee
$B$ is a constant which reflects the cost of confining the quarks inside the bag, and the $m_{q}$ are the current quark masses. 
Perturbative contributions~\cite{Kurkela:2009gj,Kurkela:2014vha}
are not included in vMIT because these become relevant at densities well above those achievable in the cores of the most massive neutron stars.

The state functions, energy density, chemical potential, and pressure, corresponding to the above Lagrangian (\textit{before} any constraints of baryon number conservation, charge neutrality, and chemical equilibrium are applied) for matter containing $u,d$ and $s$ quarks are
\beq
\ep_Q &=& \sum_{q=u,d,s}\ep_q + \frac{1}{2}a~\hbar~[\nb(y_u+y_d+y_s)]^2 
                + \frac{B}{\hbar^3}   \\
\ep_q &=& \frac{3}{8\pi^2\hbar^3}\left\{k_{Fq} (k_{Fq}^2 + m_q^2)^{1/2}(2 k_{Fq}^2 + m_q^2) 
\right.      \nn \\      
        &-& \left. m_q^4\ln\left[\frac{k_{Fq}+(k_{Fq}^2 + m_q^2)^{1/2}}{m_q}\right]\right\}   \\
\mu_q &=& (k_{Fq}^2 + m_q^2)^{1/2} + a~\hbar~\nb (y_u + y_d + y_s)        \\
P_Q &=& \nb \sum_{q=u,d,s}  \mu_q y_q - \ep_Q \,,          
\eeq
where $a \equiv (G_v/m_V)^2$ and $k_{Fq} = (\pi^2 \hbar^3 \nb y_q)^{1/3}$. The value of the vector interaction parameter $a$ is varied in the range $(0.1-0.3)~ \rm{fm}^{-2}$ to obtain different stiffness in the quark sector.

\subsection{Leptons}

Owing to the smallness of the electromagnetic fine structure constant $\alpha \simeq 1/137$, leptons are treated as non-interacting, relativistic particles for which
\beq
\ep_L &=&\frac{1}{8\pi^2\hbar^3}\sum_l\left\{k_{Fl}(k_{Fl}^2+m_l^2)^{1/2}(2k_{Fl}^2 + m_l^2)\right.  \nn \\
      &-& \left. m_l^4\ln\left[\frac{k_{Fl}+(k_{Fl}^2 + m_l^2)^{1/2}}{m_l}\right]\right\}  \\
\mu_l &=& (k_{Fl}^2 + m_l^2)^{1/2} \\
P_L &=& \nb \sum_l y_l \mu_l - \ep_L \\
k_{Fl} &=& (3\pi^2\hbar^3 \nb y_l)^{1/3} ~~;~~ l=e,\mu~. 
\eeq

At low baryon densities only electrons are present in the system, with muons appearing at a density $\nb$
such that $\mu_e -m_{\mu} = 0$. Depending on the parametrization choice, this condition also gives the density at which muons vanish.

\section{Thermodynamics of multicomponent systems}
\label{sec:multi_compt}

The original formulation of the KW approach deals with one nucleon, the neutron, and three massless quarks 
(the latter are, operationally, a single species with multiplicity 3). 
Before the KW approach can be applied to  more realistic neutron-star matter with the EOSs of the previous section, it must be generalized to include several particle species. To that end, we begin with a brief review of multicomponent thermodynamics to introduce the fundamental result from which the aforementioned generalization will be 
performed in the next section. The relations laid out below are particularly helpful in highlighting the role of the (baryon chemical potential dependent) switch function $S$ which is instrumental in realizing a crossover transition.

The number density of a single-component system in the grand-canonical ensemble is given by the total derivative of the pressure with respect to the chemical potential, 
\be
n=\frac{dP}{d\mu}~.
\ee
The equivalent expression for a multicomponent system is obtained from the grand potential
\be 
\Phi(T,V,\mu_i)=U-TS-\sum_i N_i\mu_i
\ee
or, in units of energy-density, 
\be
\phi=\ep-Ts-\sum_i n_i\mu_i ~.
\ee
The differential of $\phi$ is
\be
d\phi=d\ep-sdT-\sum_i n_i d\mu_i
\ee
which implies that the number density of particle species $i$ is given by 
\be
n_i=-\left.\frac{\p \phi}{\p \mu_i}\right|_{T,\mu_j}~.
\ee
The thermodynamic identity 
\be
\ep = Ts-P + \sum_i n_i\mu_i
\label{TI}
\ee
means that $P=-\phi$ and therefore  
\be
n_i=\left.\frac{\p P}{\p \mu_i}\right|_{\mu_j}  \label{eqni}
\ee
where the temperature $T$ has been suppressed as, in what follows, only cold matter is considered.

\Eqn{eqni} is central to the subsequent discussion where we show the manner in which the individual number densities of the various nucleonic and quark species are modified by the switch function $S$ of the KW machinery.

\section{Unconstrained and beta-equilibrated crossover matter}
\label{sec:KWG}

In this section, we start with the crossover EOS where 
baryon number conservation, charge neutrality, and weak interaction equilibrium are not imposed, i.e., ``unconstrained'' matter. Working with  unconstrained quantities enables us to calculate the various partial derivatives required in the determination of the squared adiabatic speed of sound $\csqad$ (see Sec.~\ref{sec:soundspeed}) prior to the imposition of the conditions mentioned above.

In the KW description of crossover matter, the pressure is given by
\be
P_B = (1-S)P_H + S \, P_Q     \label{eqPB}
\ee
where $P_H$ and $P_Q$ are the hadron and quark pure-phase pressures respectively, and the switch function
\be
S = \exp\left[-\left(\frac{\mu_0}{\mu}\right)^4\right]
\ee
gives the fraction of quark matter to the total baryonic matter in the crossover setting, with $\mu$ being the average hadronic chemical potential
\be
\mu = \frac{n_n \mu_n + n_p \mu_p}{n_n + n_p}~\,,
\ee
and $\mu_0$ a typical energy scale for the crossover. This choice for $\mu$ will be justified in the next section.

Applying \Eqn{eqni} to hadrons leads to
\beq
n_h^* &=& (1-S)\frac{\p P_H}{\p \mu_h}+S\frac{\p P_Q}{\p \mu_h}+(P_Q-P_H)\frac{\p S}{\p \mu_h} \nn \\
&=& (1-S)n_h+0+(P_Q-P_H) \frac{4\mu_0^4S}{\mu^5}\frac{\p \mu}{\p \mu_h} \nn \\
&=& (1-S)n_h+(P_Q-P_H) \frac{4\mu_0^4S}{\mu^5}\frac{n_h}{n_n+n_p}   \nn  \\
&=& n_h\left[1-S\left(1-\frac{4\mu_0^4}{\mu^5}\frac{P_Q-P_H}{n_n+n_p}\right)\right] \,,
\label{eqnh}
\eeq
where $n_i^*=\nb y_i^*$ refers to a crossover-matter density and $n_i = \nb y_i$ to a pure-phase density. Thus, in the present context, the starred fractions are the physical quantities whereas the unstarred ones are merely bookkeeping devices. For leptons this distinction is irrelevant. 

For quarks, one obtains
\beq
n_q^* &=& (1-S)\frac{\p P_H}{\p \mu_q}+S\frac{\p P_Q}{\p \mu_q}+(P_Q-P_H)\frac{\p S}{\p \mu_q} \nn \\
&=& 0+S\,n_q+0 = S\,n_q \,. 
\label{qno}
\eeq

Finally, the energy density $\ep$ is obtained from \Eqn{TI} using \Eqn{eqPB} for the pressure, Eqs.~(\ref{eqnh})-(\ref{qno}) for the number densities of hadrons and quarks, respectively, and the pure-phase chemical potentials defined in Sec.~\ref{sec:EOS}.

\subsection{Beta-equilibrium}

We turn now to the discussion of neutron-star matter that consists of nucleons, leptons and quarks.
Initially, the system is entirely unconstrained with $\nb$ and $y_i$ ($i=n,p,u,d,s,e,\mu$) as the free variables. Then, strong equilibrium
\be
\mu_n = 2\mu_d + \mu_u  ~~;~~\mu_p = 2\mu_u + \mu_d 
\ee
and weak equilibrium
\be
\mu_n = \mu_p + \mu_e  ~~;~~\mu_e = \mu_\mu   ~~;~~ \mu_d = \mu_s
\ee
are enforced, as well as charge neutrality
\be
n_p^*+(2n_u^*-n_d^*-n_s^*)/3-(n_e+n_{\mu})=0 
\ee
and baryon number conservation
\be
n_n^*+n_p^*+(n_u^*+n_d^*+n_s^*)/3-\nb=0 ~.
\label{barno}
\ee
These conditions eliminate the particle fractions in favor of the total baryon density: 
\be
y_i \rightarrow y_{i,\beta}(\nb)  ~~;~~ i=n,p,u,d,s,e,\mu
\ee

\subsection{Comparison of Kapusta-Welle (KW) with McLerran-Reddy (MR) and Zhao-Lattimer (ZL) EOSs}

In this subsection, we briefly discuss interesting similarities and differences between the crossover model of KW~\cite{Kapusta:2021ney} and recently proposed quarkyonic ``shell'' models of MR~\cite{McLerran:2018hbz} and ZL~\cite{Zhao:2020dvu}, and explain the reason why the latter is not suitable for \gm~calculations in its present form.
The baryon number densities in the quarkyonic matter descriptions of MR~\cite{McLerran:2018hbz} and ZL~\cite{Zhao:2020dvu} are 
\beq
n_h^* &=& \frac{k_{Fh}^3-k_{0h}^3}{3\pi^2\hbar^3}
=\frac{k_{Fh}^3}{3\pi^2\hbar^3}\left(1-\frac{k_{0h}^3}{k_{Fh}^3}\right)
= n_h\left(1-\frac{k_{0h}^3}{k_{Fh}^3}\right) \,, \nn \\
\eeq
where $k_{0h}$ are the minimum momenta of hadrons or nucleons 
in quarkyonic matter, which depend on the corresponding Fermi momenta $k_{Fh}$ and thus the baryon number density. The precise way in which $k_{Fh} - k_{0h}$ depends on a chosen momentum scale $\Lambda$ and a common transition density $n_t$ is detailed in Eq.~(17) of Ref.~\cite{Zhao:2020dvu}.

Comparing the above expression to \Eqn{eqnh} from the previous section, it becomes clear that the presence of the hadron shell in the MR and ZL approaches forces hadrons to higher-momentum states much like $S$ does in the KW scheme:
\be
S\left(1-\frac{4\mu_0^4}{\mu^5}\frac{P_Q-P_H}{n_n+n_p}\right) \stackrel{\wedge}{=} \frac{k_{0h}^3}{k_{Fh}^3} \,.
\ee
This means that any particle species participating in $S(\mu)$ (that is, a species $i$ for which $\p \mu/\p \mu_i\neq0$) will invariably inherit a shell-like term in its  crossover-matter number density. 
In quarkyonic matter realizations,  such a term is desirable for baryons but not for quarks and therefore $\mu$ must be a function of baryonic chemical potentials only. 

Note that in the case of KW, the quark densities in crossover-matter $n_q^*$ are the product of the corresponding pure-phase densities and the quark-to-baryon fraction $S$ which is an \textit{a priori} assumption. 
On the other hand, for both MR and ZL models, the densities and fractions of baryons and quarks in the quarkyonic phase are determined by the solution of the equilibrium equations. 

Here, we should point out that in the MR and ZL implementations of the quarkyonic matter scenario, the nucleonic Fermi momenta are weakly dependent on baryon density when the latter exceeds the transition density, $n_t$; that is, 
\be
|k_{\infty,i}-k_{Fi,\beta}|/k_{\infty,i}\ll 1 ~~\text{for all}~ \nb>n_t \,,
\ee
where $k_{\infty,i}\equiv k_{Fi,\beta}(\nb \rightarrow \infty)$.
On the other hand, the nucleonic chemical potentials and, by extension, the pressure change very rapidly with $k_{Fi}$ for $\nb>n_t$ due to the presence of denominators $\propto(1-K_i)$ (see Eqs.~(19)-(20) in \cite{Zhao:2020dvu}) in their kinetic parts, where
\beq
K_i^{\rm MR} &=& \left(\frac{k_{0i}}{k_{Fi}}\right)^2 \left(1 + \frac{2\Lambda^3}{k_{Fi}^3}\right)\\
K_i^{\rm ZL} &=& \left(\frac{k_{0i}}{k_{Fi}}\right)^2 \left(1 + \frac{\Lambda^2}{k_{Fi}^2}\right)~.
\eeq
These terms remain close to zero [i.e. $(1-K_i)^{-1}\rightarrow \infty$], in equilibrium matter as a result of the aforementioned behavior of the nucleonic Fermi momenta, throughout the quarkyonic regime. 
These two features of MR and ZL are responsible for divergent pressure derivatives with respect to $k_{Fi}$ which, in turn, lead to superluminal adiabatic sound speeds. Thus MR and ZL, in their current formulations, are unsuitable for our purposes and we do not apply them in \gm~calculations for crossover matter.

\section{Sound speeds}
\label{sec:soundspeed}

In this section, we describe how calculations of the squared adiabatic and equilibrium sound speeds, $\csqad$ and $\csqeq$, required in the calculation 
of \gm~frequencies, are performed. As one of our objectives is to provide contrasts between \gm~frequencies in crossover matter and the case of a first-order transition treated via the Gibbs construction, both cases are considered below.

\subsection{Sound speeds in crossover matter}

Within the KW framework, the total pressure and energy density in the crossover region are 
\beq
P &=& P_B + P_e + P_{\mu}  \\
\ep &=& \ep_B + \ep_e + \ep_{\mu}  \\
\ep_B &=& -P_B + \sum_{i=n,p,u,d,s}n_i^*\mu_i  \,.
\eeq

Using these, the adiabatic speed of sound is obtained by first calculating the expression
\be
\csqad(\nb,y_i) = \left.\frac{\p P}{\p \nb}\right|_{y_i}
\left(\left.\frac{\p \ep}{\p \nb}\right|_{y_i}\right)^{-1}
\ee
and then evaluating it in $\beta$-equilibrium
\be
c_{\rm{ad},\beta}^2(\nb) = \csqad[\nb,y_{i,\beta}(\nb)]~.
\ee
The equilibrium sound speed is given by the total derivatives of the pressure and the energy density  with respect to the baryon density after the enforcement of $\beta$-equilibrium,
\be
\csqeq = \frac{dP_{\beta}}{d\nb}\left(\frac{d\ep_{\beta}}{d\nb}\right)^{-1}~.
\ee
%

\subsection{Sound speeds with Gibbs construction}

As in the crossover matter case, all thermodynamic quantities are expressed in terms of functions of the total baryon density $\nb$, and the individual particle fractions $y_n$,~$y_p$,~$y_e$,~$y_{\mu}$,~$y_u$,~$y_d$,~$y_s$ which are, at this point, independent variables. That is, 
\beq
\ep_H &=& \ep_H(\nb,y_n,y_p) ~;~ P_H = P_H(\nb,y_n,y_p) ~; \nn \\
\mu_h &=& \mu_h(\nb,y_n,y_p)  \\ \nn \\
\ep_Q &=& \ep_Q(\nb,y_u,y_d,y_s) ~;~ P_Q = P_Q(\nb,y_u,y_d,y_s) ~; \nn \\
\mu_q &=& \mu_q(\nb,y_q) \\ \nn \\
\ep_L &=& \ep_L(\nb,y_e,y_{\mu}) ~;~ P_L = P_L(\nb,y_e,y_{\mu}) ~; \nn \\
\mu_l &=& \mu_l(\nb,y_l)  \,.
\eeq

The conditions for weak equilibrium, charge neutrality, and baryon number conservation are applied afterwards.  
These introduce another independent variable, $\chi$, which is the volume fraction of quarks in the mixed phase of Gibbs construction:
\beq
&&P_H = P_Q ~;~ \mu_n = 2\mu_d + \mu_u ~;~ \mu_p = 2\mu_u + \mu_d  \\ \nn \\
&&\mu_n = \mu_p + \mu_e ~;~ \mu_e = \mu_\mu ~;~\mu_d = \mu_s  \\ \nn \\
&&3 (1-\chi) y_p+\chi(2y_u-y_d-y_s)-3(y_e+y_{\mu}) = 0   \\  \nn \\
&&3 (1-\chi)(y_n+y_p)+\chi(y_u+y_d+y_s)-3 = 0 \,.
\eeq 

Solving these equations eliminates the $y_i$ and $\chi$ in favor of $\nb$. Thus the state variables become functions of only $\nb$ according to the rule
\[ Q(\nb,y_i,y_j,...) \rightarrow Q[\nb,y_i(\nb),y_j(\nb),...] = Q(\nb)~.\] \\
Then, the thermodynamics of the mixed $(^*)$ phase are:
\beq
\ep^* &=& (1-\chi)\ep_H + \chi \ep_Q + \ep_L  \\
P^* &=& P_H+ P_L = P_Q+ P_L \nn \\
    &=& (1-\chi) P_H + \chi P_Q + P_L  \\
\mu_h^* &=& \mu_h ~~;~~ \mu_q^* = \mu_q   \\
y_h^* &=& (1-\chi)y_h  ~~;~~ y_q^* = \chi y_q \,.
\label{qnoG}
\eeq
Quantities corresponding to leptons are not affected by the ratio of the two baryonic components in the mixed phase.

\begin{table}[h]
\caption{Parameter sets for the EOSs used in this work. } 
\begin{center} 
\begin{tabular}{ccrrrc}
\hline
\hline
Model       & Parameter  & XOA     & XOB     & XOC  & Units \\ \hline
            &  $a_0$     & -96.64   & -90.39   & -96.64 & MeV     \\
            &  $b_0$     &  58.85   &  52.60   & 58.85  & MeV     \\
  ZL        &  $\gamma$  &  1.40    &  1.446   & 1.40   &         \\
            &  $a_1$     & -26.06   & -232.78  & -28.15 & MeV     \\
            &  $b_1$     &  7.34    &  212.46  & 7.83   & MeV     \\
            & $\gamma_1$ &  2.45     &  1.1     & 3.5    &         \\
\hline 
            & $m_u$      &  5.0     &  5.0     & 5.0    & MeV     \\
            & $m_d$      &  7.0     &  7.0     & 7.0    & MeV     \\
  vMIT      & $m_s$      &  150.0   &  150.0   & 150.0  & MeV     \\
            & $a$        & 0.20     &  0.23    & 0.15   & fm$^2$  \\
            & $B^{1/4}$  & 180.0    &  180.0   & 180.0  & MeV     \\
\hline
  KW        & $\mu_0$    & 1.8      &  1.8     & 1.8    & GeV     \\
\hline \hline
\end{tabular}
\end{center}
\label{tab:Parameters}
\end{table}

The mixed phase extends over those densities $\nb$ for which $0\le \chi(\nb) \le 1$. In contrast to the crossover case where $S$ operates at \textit{all} densities, $\chi$ is active only when the above condition is satisfied. 
Moreover, 
in the case of a 
Gibbs construction of the first-order phase transition scenario, $\chi$ and $y_i$ are treated on an equal footing with no prior assumptions regarding their density dependence, whereas in the crossover 
scenario, $S$ has a definitive functional form which the particle fractions must be adjusted to fit. 
As a consequence, even though 
both $\chi$ and $S$  
describe the quark-to-baryon fraction, the former is a quantity for which we solve while the latter acts as a constraint replacing the Gibbs condition for mechanical equilibrium.

The sound speeds are obtained following the prescription outlined in the previous subsection. Alternatively, the adiabatic sound speed in the mixed phase can be calculated from the corresponding ones in the pure hadronic and quark phases separately according to Ref.~\cite{Jaikumar:2021jbw}
\be
\frac{1}{\cad^{*2}} = \frac{1-\chi}{c_{{\rm{ad}},H}^2} + \frac{\chi}{c_{{\rm{ad}},Q}^2}~.
\ee 
%

\section{Results}
\label{sec:Results}

We turn now to present results based on calculations of the crossover EOS, associated NS properties, the two sound speeds and the resulting \gm~frequencies. For contrast, results corresponding to pure hadronic matter and those for a first-order phase transition treated using the Gibbs construction are also presented.

\subsection{EOS and structural properties of NSs}

To construct crossover models, we have chosen the parameter values shown in Table~\ref{tab:Parameters}, labeled as XOA, XOB and XOC, for the parametrization of the EOSs used in this work.~\footnote{The method to determine the constants for the ZL parametrization is described in Refs.~\cite{Zhao:2020dvu} and \cite{Jaikumar:2021jbw}.} 
These sets of parameters correspond to the nuclear and neutron-star properties shown in Table~\ref{tab:Properties}.

\begin{table}[h]
\caption{Nuclear and neutron-star properties corresponding to the parametrizations shown in Table~\ref{tab:Parameters}. 
The symbols refer to 
$\nsat$: nuclear saturation density, 
$E_0$: energy per particle at $\nsat$, 
$K_0$: compression modulus of symmetric nuclear matter (SNM) at $\nsat$, 
$S_v$: symmetry energy at $\nsat$, 
$L$: slope of $S(n)$ at $\nsat$, 
$n_{\mu,\text{on}}^{\beta}$: onset density of muons, 
$n_{\mu,\text{off}}^{\beta}$: turnoff density of muons, and $n_{p,\text{off}}^{\beta}$: turnoff density of protons. 
Quantities related to neutron stars are 
$R$: radius, $M$: mass, $\beta=GM/Rc^2$: compactness, 
$n_c$: central density, $p_c$: central pressure, 
$\ep_c$: central energy density, 
$\Lambda$: dimensionless tidal deformability, 
$\csqeq$: squared equilibrium sound speed, 
and $\csqad$: squared adiabatic sound speed. 
The subscripts $1.4$ and ${\rm max}$ denote the masses of stars in $\Msolar$.}    
\begin{center} 
\begin{tabular}{crrrrrc}
\hline
\hline
Property      &  XOA  &  XOB  &  XOC  &   ZL    &  Gibbs  & Units \\ \hline
$\nsat$         & 0.16   & 0.16    &  0.16   &   0.16  &  0.16   & fm$^{-3}$  \\
$E_0$     & -16.0  & -16.0   &  -16.0  &  -16.0  &  -16.0  & MeV  \\ 
$K_0$         & 250.0  & 260.0   &  250.0  &  250.0  &  250.0  & MeV \\
$S_v$         & 31.6   & 30.0    &  30.0   &  31.6   &  31.6   & MeV \\
$L$           & 43.0   & 70.0    &  65.0   &  43.0     &  43.0   & MeV \\ 
$n_{\mu,\text{on}}^{\beta}$  & 0.13 & 0.15 & 0.14      &  0.13   &  0.13    & fm$^{-3}$ \\
$n_{\mu,\text{off}}^{\beta}$ & 1.39 & 0.77 & 1.67      &  N/A    &  1.32     &fm$^{-3}$  \\
$n_{p,\text{off}}^{\beta}$   & N/A & 0.88  &  N/A      &  N/A    &  N/A     &fm$^{-3}$  \\
\hline 
$\Rtyp$             & 12.4   & 12.4    &  13.8    &    12.4  &    12.4  &{\rm km} \\
$\beta_{1.4}$         & 0.167  & 0.166   &  0.150   &    0.166 &    0.166 & \\
$n_{c,1.4}/\nsat$       & 2.64   & 3.16    & 1.96     &    2.64  &    2.68  & \\
$p_{c,1.4}$           & 60.2   & 73.6    & 41.7     &   60.0    &  69.8  & ${\rm MeV~fm^{-3}}$ \\
$\ep_{c,1.4}$       & 424.7  & 518.4   &  316.7   &  424.9   &  436.3 & ${\rm MeV~fm^{-3}}$ \\
$\Lambda_{1.4}$       & 428.9  & 426.7   &  841.3   &  430.1   &  421.2 & \\
$(\csqeq)_{c,1.4}$  & 0.398  & 0.331   &  0.294   &  0.397   &  0.406 & $c^2$ \\
$(\csqad)_{c,1.4}$  & 0.429  & 0.331   &  0.497   &  0.426   &  0.528 & $c^2$ \\
\hline
$\Rmax$             & 11.5   & 10.4   &  13.0     &  11.1      &  11.2     &{\rm km} \\
$\Mmax$             & 2.11   & 2.04   &  2.13     &  2.23      &  2.08     &$\Msolar$ \\ 
$\beta_{\rm max}$         & 0.270  & 0.289  &  0.242    &  0.295     &  0.275    &     \\
$n_{\rm c,max}/\nsat$       & 5.83   & 7.38   &  4.70     &  6.14      &  6.32     &     \\
$p_{\rm c,max}$           & 362.7  & 696.1  &  213.4    &  577.0     &  457.7    & ${\rm MeV~fm^{-3}}$ \\
$\ep_{\rm c,max}$ & 1142.9 & 1549.5 &  886.8    &  1202.4    & 1264.9    & ${\rm MeV~fm^{-3}}$ \\
$\Lambda_{\rm max}$       & 13.8   & 6.4    &  30.1     &  6.2       & 11.1      &   \\
$(\csqeq)_{\rm c,max}$  & 0.426  & 0.566  &  0.316    &  0.767     & 0.576     & $c^2$ \\
$(\csqad)_{\rm c,max}$  & 0.507  & 0.818  &  0.353    &  0.889     & 0.653     & $c^2$ \\
\hline \hline
\end{tabular}
\end{center}
\label{tab:Properties}
\end{table}

For XOA, all values are within 1-$\sigma$ deviations of empirical/observational constraints discussed below. While XOB and XOC do not perform as well, they are used here to illustrate some important physics related to the behavior of the \gm~frequency.
Specifically, with XOB we investigate \gm~frequency features corresponding to a sound-speed peak due to proton disappearance, whereas in XOC the peak in the speed of sound is not related to a change in the number of degrees of freedom.
Models labeled ZL (nucleons only) and Gibbs (nucleons plus quarks with a Gibbs construction) use the appropriate parameters of XOA.
\vspace{3mm}

\begin{figure}[h!]
\includegraphics[scale=0.32]{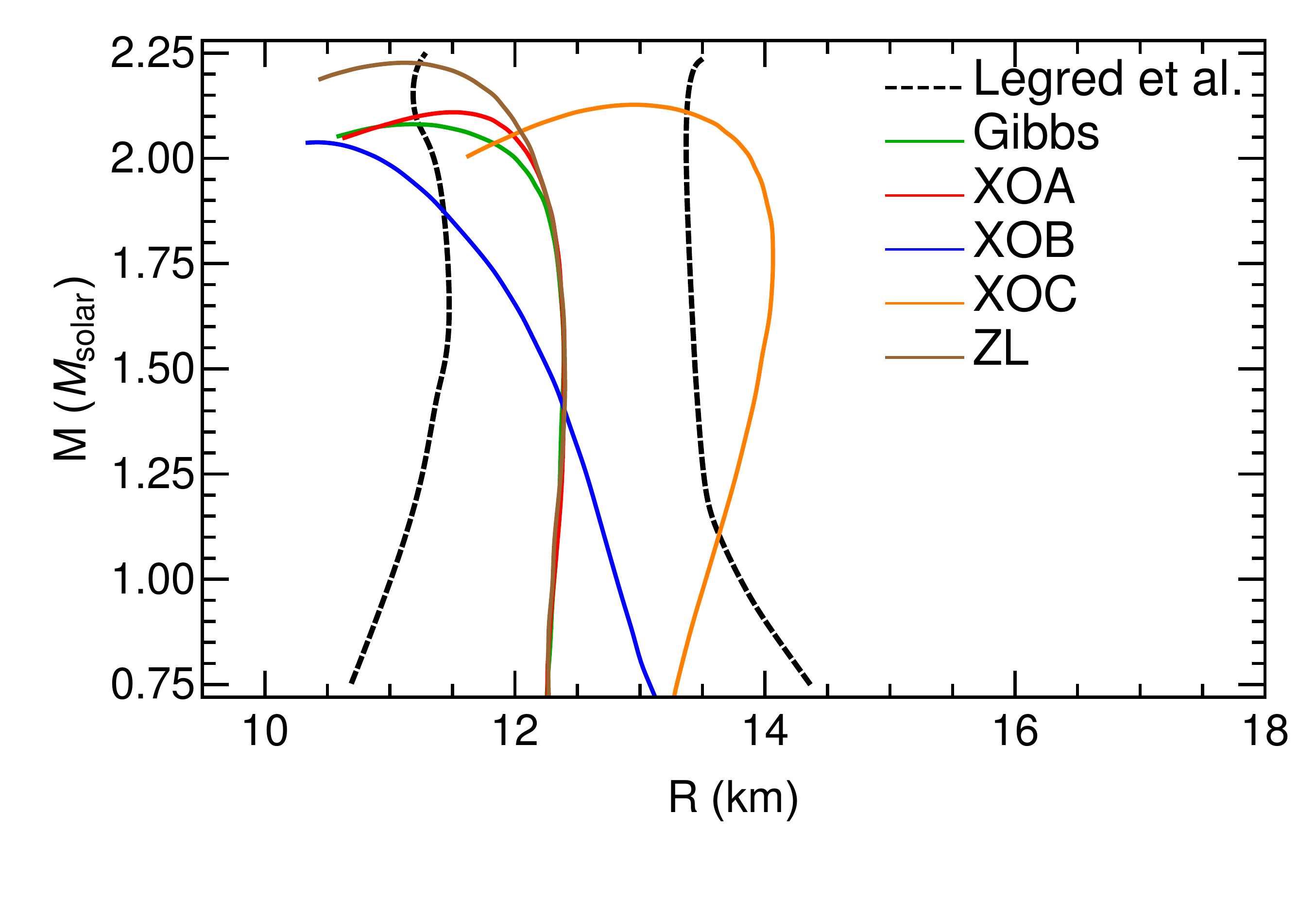}
\caption{Neutron-star $M$-$R$ curves for the various EOS models used in this work. The black, dashed lines represent the 90\%-confidence level constraints extracted from recent radio, x-ray, and gravitational-wave observations by Legred et al.~\cite{Legred:2021hdx}. 
Models corresponding to the ``A'' parameter set fit these constraints well with differences between the three depending on the order of the transition to quark matter or the lack thereof. 
While stars with $M\leq 1.8\,\Msolar$ using model XOB are within the constraints, with model XOC only stars close to the maximum mass satisfy the Legred et al. constraints.
} 
\label{fig:MR}
\end{figure}

\begin{figure*}[htb]
\parbox{0.5\hsize}{
\includegraphics[width=\hsize]{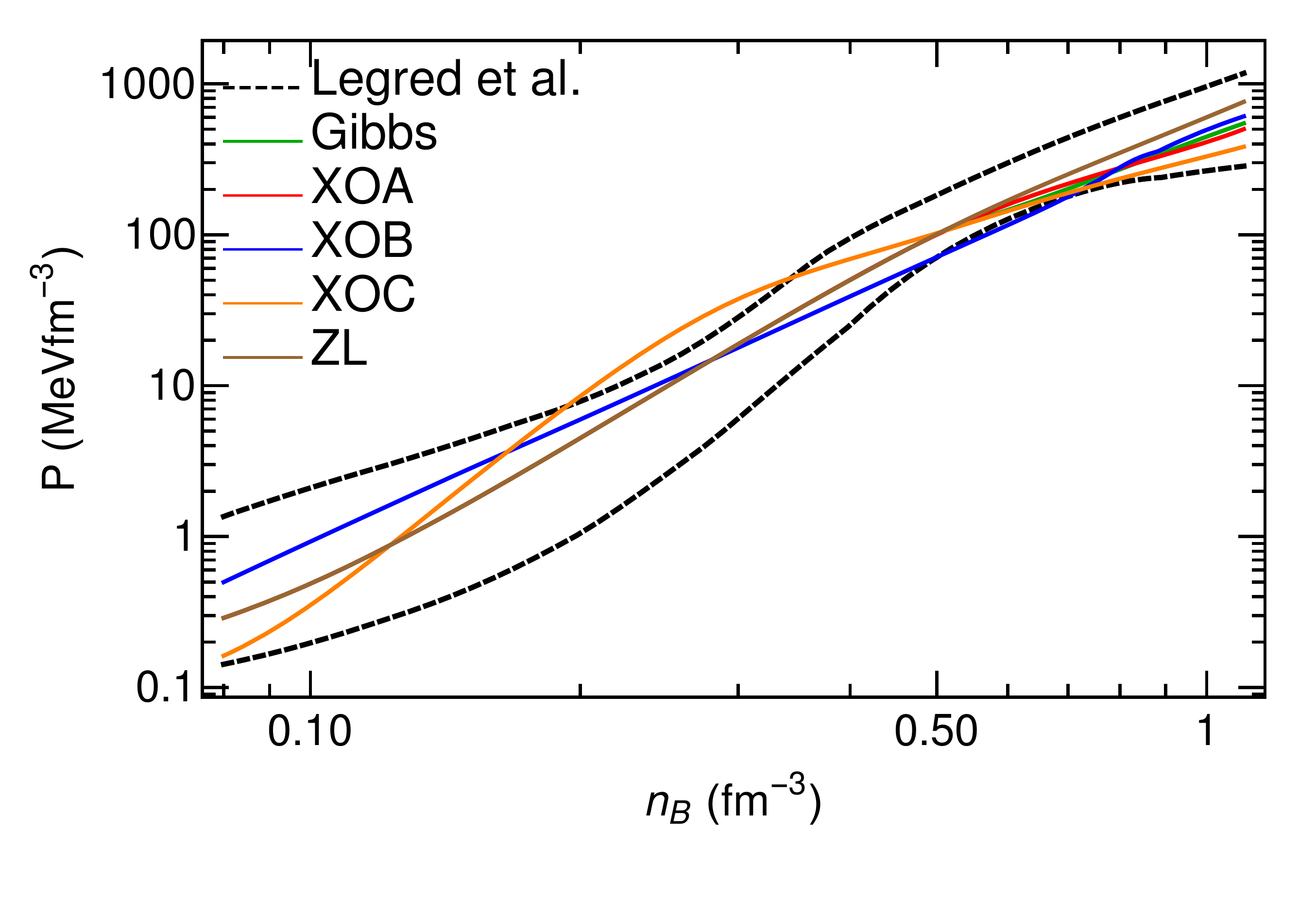}\\[-4ex]
}\parbox{0.5\hsize}{
\includegraphics[width=\hsize]{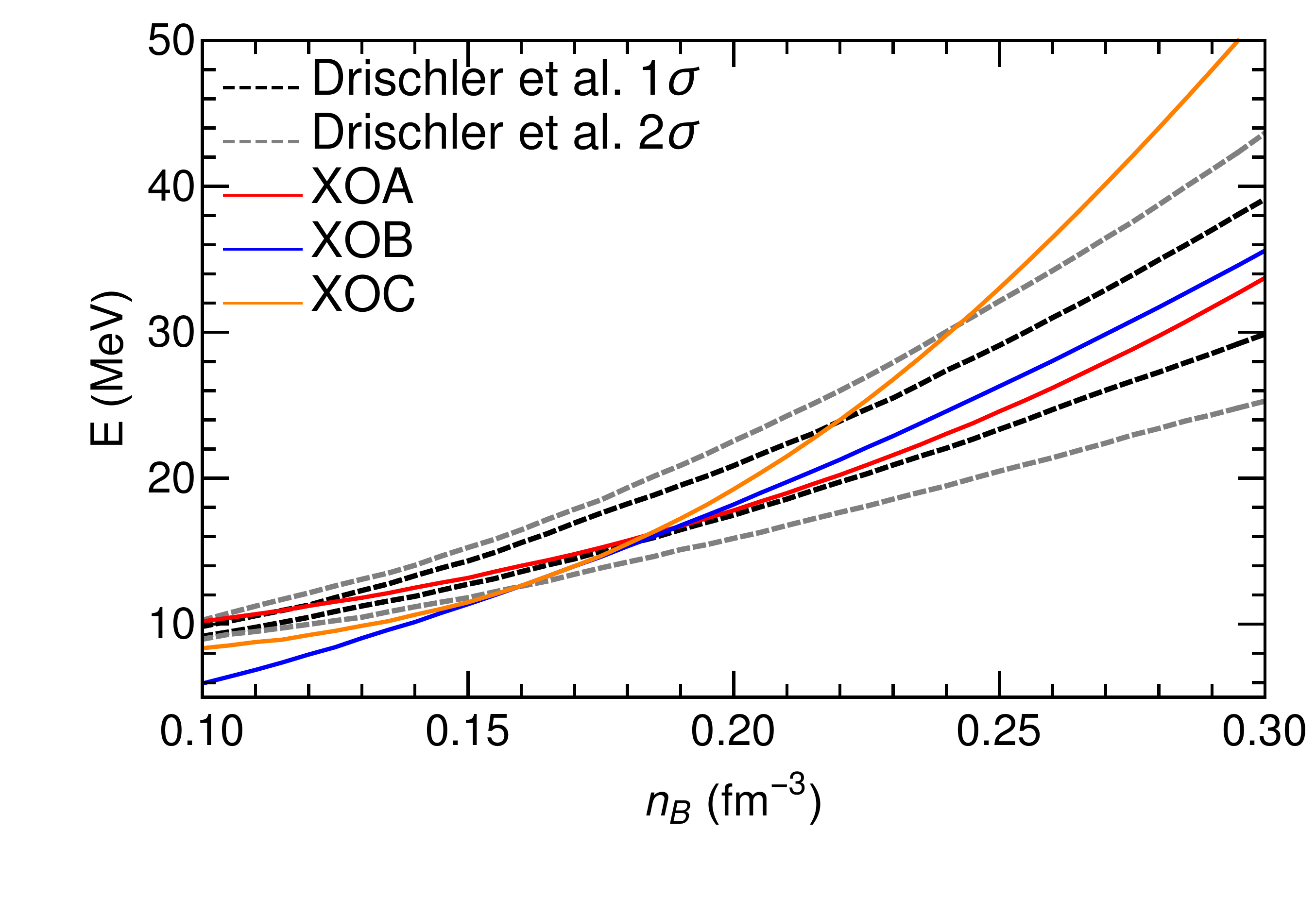}\\[-4ex]
}
\caption{Left panel: pressure versus baryon density as obtained by the assorted EOSs used herein compared with the astrophysical constraints of Legred et al.~\cite{Legred:2021hdx} (in black, dashed lines). 
All parametrizations meet these constraints successfully with the exception of XOC which fails to stay in the allowed region around $2.0\,\nsat$, where $\nsat=0.16~{\rm fm}^{-3}$. 
Right panel: energy-per-particle versus baryon density of beta-equilibrium matter for the various EOSs used in this work compared with the 1-$\sigma$ and 2-$\sigma$ constraints 
from Drischler et al.~\cite{Drischler:2020fvz} obtained in a chiral EFT framework. Results corresponding to Gibbs and ZL are not shown because, over the range of densities displayed, they are identical to XOA. 
Only XOA remains within the 2-$\sigma$ constraints of \cite{Drischler:2020fvz} up to $\sim2.0\,\nsat$.
}
\label{fig:PN_EN_nB}
\end{figure*}

We wish to note that the values of the symmetry energy $S_v$ and and its slope $L$ at $\nsat$ used in our work (see Table~\ref{tab:Properties}) lie in the range $\simeq 31 \pm 2$ MeV and $\simeq 51\pm 11$ MeV, respectively, recommended in Ref.~\cite{Lattimer:2012xj}. These values led to the bounds on the radius of a $1.4\,\Msolar$ star to be $\Rtyp \simeq 12 \pm 1$ km.  
\vspace{3mm}

Interpretations of the recent PREX-II experiment carried out at the Jefferson Lab (JLab) measuring the neutron skin thickness of $^{208}$Pb~\cite{Adhikari:2021phr}, $\Rskin = 0.283 \pm 0.071$~fm, however, widely vary in their inferences of the appropriate values of $S_v$ and $L$ to be used. 
For example, Reed et al.~\cite{Reed:2021nqk}, using relativistic mean-field theory (RMFT) calculations to analyze the JLab data, conclude that $S_v=38.1\pm 4.7$ MeV and $L=106\pm 37$ MeV, values that are significantly higher than those deduced in earlier works. Furthermore, the bound $\Rtyp > 13.25$~km was found there.  
Reinhard et al.~\cite{Reinhard:2021utv}, use covariant RMFT (with density-dependent couplings) and nonrelativistic energy functionals to analyze the PREX-II data and 
combine it with the dipole polarizability data of $^{208}$Pb 
to arrive at $S_v = 32\pm 1$ MeV and $L=54\pm 8$ MeV. In addition, these authors obtain $\Rskin = 0.19\pm 0.02$~fm in accord with earlier deductions. 
Similar results are obtained by Essick et al.~\cite{Essick:2021kjb} who report $S_v = 34\pm 3$ MeV, $L=58\pm 18$ MeV and $\Rskin = 0.19~^{+0.03}_{-0.04} $~fm from a non-parametric EOS coupled with Gaussian processes.
Combining recent mass and radius measurement from radio and x-ray data from \textit{NICER}, Biswas~\cite{Biswas:2021yge} finds $\Rskin = 0.20\pm 0.05$~fm and $\Rtyp = 12.75~^{+0.42}_{-0.54}$ km using nuclear EOSs with piecewise polytrope parametrization. 
Given the fluid state of theoretical inferences from the analysis of JLab data, we have opted to stick with the values used in Table~\ref{tab:Properties}.  
\vspace{3mm}

Figure~\ref{fig:MR} shows mass versus radius ($M$-$R$) curves for all the models considered along with the 
recent constraints obtained by Legred et al.~\cite{Legred:2021hdx}, which combined available observations including the radio mass measurements of PSR J0348+0432 and J0470+6620~\cite{Fonseca:2021wxt,Cromartie:2019kug,Antoniadis:2013pzd}, the mass and tidal deformability measurements of GW170817 and GW190425~\cite{Abbott:2018wiz,LIGO:2017qsa,Abbott:2020uma}, and the x-ray mass and radius constraints from latest \textit{NICER} measurements of J0030+0451 and J0470+6620~\cite{Miller:2019cac,Riley:2019yda,Miller:2021qha,Riley:2021pdl}.  
The constraints of Legred et al. were obtained by using the hierarchical inference~\cite{Loredo:2004nn} and a nonparametric survey through Gaussian Processes (GPs) conditioned on existing EOS models in the literature~\cite{Landry:2020vaw,Landry:2018prl,Essick:2019ldf}. 
\vspace{3mm}

The left panel of Fig.~\ref{fig:PN_EN_nB} displays results of the pressure versus baryon number density ($P-\nb$) relation for the various models used in the present work, in contrast to those inferred from Ref.~\cite{Legred:2021hdx} mentioned above (the black dashed boundaries, adapted from their Fig.~4). 
To provide a comparison, results of energy versus density $E$ vs $\nb$ of $\beta$-equilibrated neutron-star matter (NSM) are shown in the right panel of Fig.~\ref{fig:PN_EN_nB}, 
together with those for NSM from the chiral effective theory calculations of Ref.~\cite{Drischler:2020fvz} where 1-$\sigma$ and 2-$\sigma$ error estimates up to $\sim2.0\,\nsat$ were provided. 
Although not shown, we also find that results of the crossover, ZL and Gibbs models for $P-\nb$ and $E$ versus $\nb$ are consistent with microscopic Greens' function calculations of Gandolfi et al.~\cite{Gandolfi:2011xu}. 
\vspace{3mm}

\begin{figure*}[htb]
\parbox{0.5\hsize}{
\includegraphics[width=\hsize]{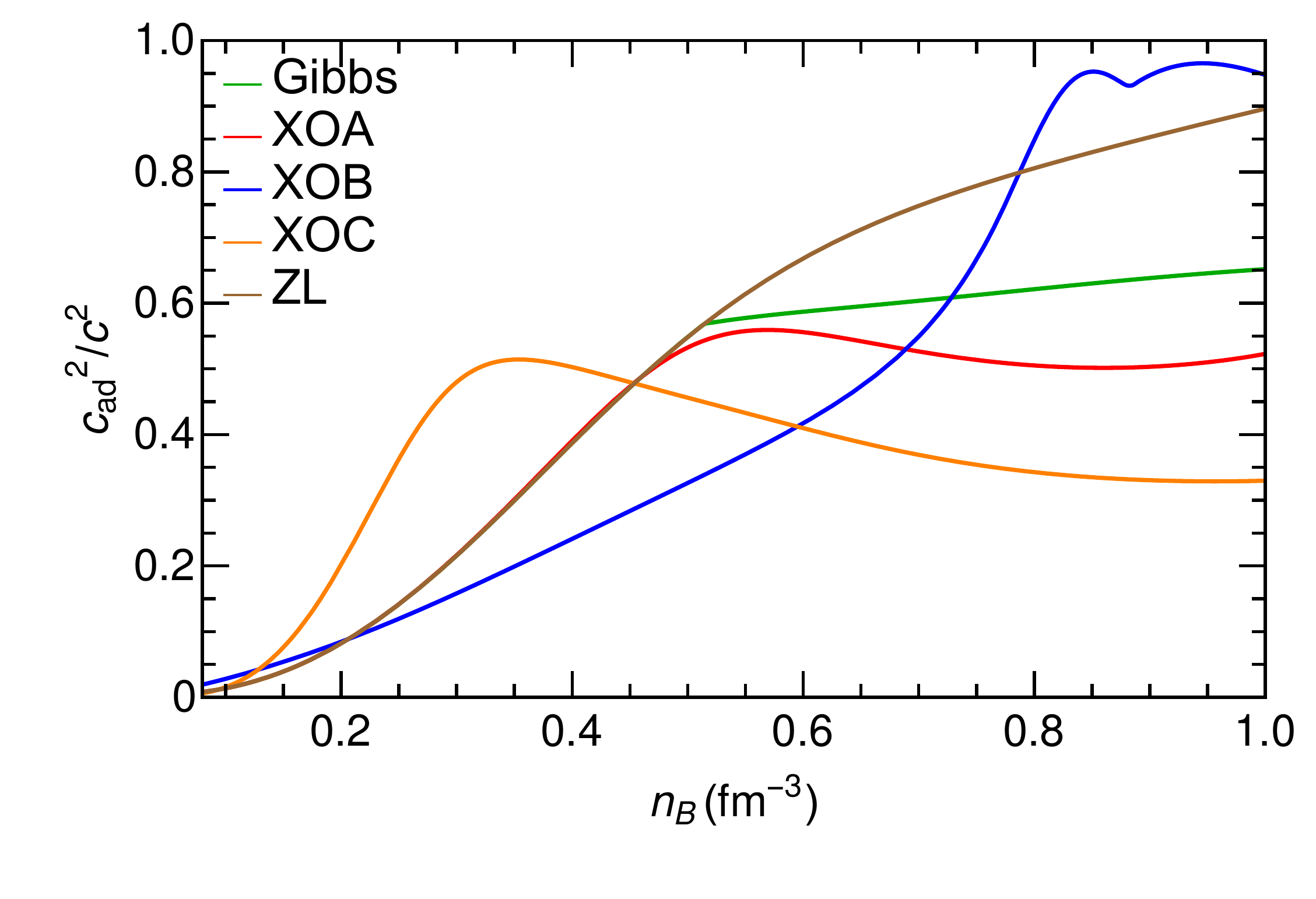}\\[-4ex]
}\parbox{0.5\hsize}{
\includegraphics[width=\hsize]{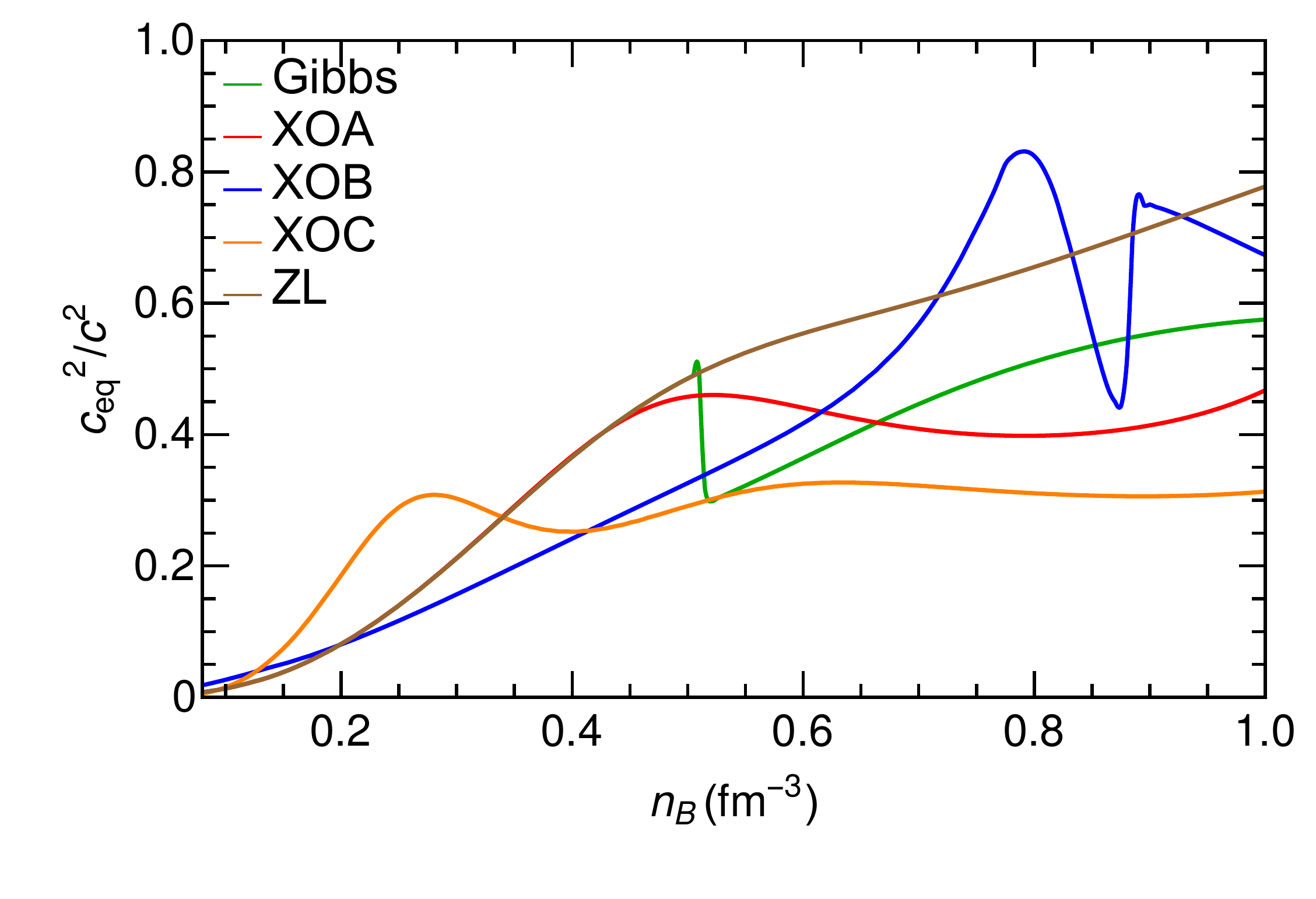}\\[-4ex]
}
\caption{Left panel: the squared adiabatic sound speed $\csqad$ as a function of the baryon density $\nb$. 
Right panel: the squared equilibrium sound speed $\csqeq$ as a function of the baryon density $\nb$.
}
\label{fig:cs_ad_eq}
\end{figure*}

The squared adiabatic and equilibrium sound speeds $\csqad$ and $\csqeq$ versus baryon density $\nb$ are shown in the left and right panels of Fig.~\ref{fig:cs_ad_eq}, respectively. Both $\csqad$ and $\csqeq$ increase monotonically with $\nb$ for the ZL model in which nucleons are the only baryons. The non-monotonic behaviors of the other curves are due to admixtures of nucleons and quarks in the baryon sector. The $\csqeq (\nb)$ for the Gibbs model suddenly drops (rises) at the onset (end) of the mixed phase (the latter not shown in the figure), whereas $\csqad (\nb)$ varies smoothly. Results for the crossover models XOA and XOC are similar in structure, whereas those for XOB show more structure at large $\nb$ due to the disappearance of protons. With the exception of model XOB, $\csqad > \csqeq$ for all other models at all $\nb$.
\vspace{3mm}

\begin{figure}[h!]
\includegraphics[scale=0.38]{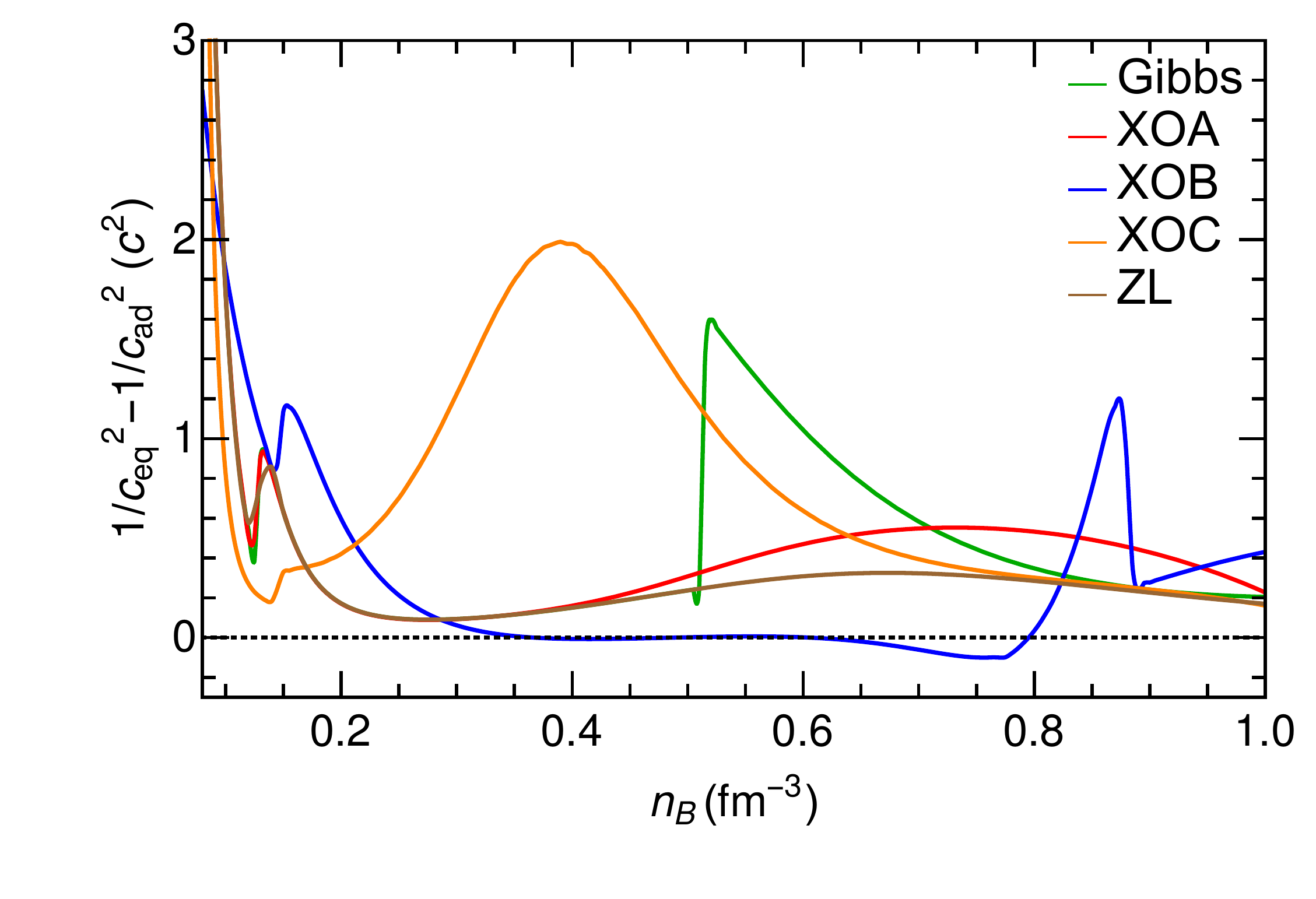}
\caption{Difference between the inverse-squared sound speeds versus the baryon number density. 
The peaks in the vicinity of $\nsat$ correspond to muon appearance and are present in all models. 
The peak at $\sim3\,\nsat$ for Gibbs occurs at the onset of the mixed phase which extends beyond the densities shown here. 
The peak around $5.5\,\nsat$ for XOB is the combined effect of muon and proton disappearance in this model. 
On the other hand, the peak at $\sim2.5\,\nsat$ for XOC results from inflection points in the quark and neutron fractions.}
\label{fig:OOC}
\end{figure}

\begin{figure}[h!]
\includegraphics[scale=0.32]{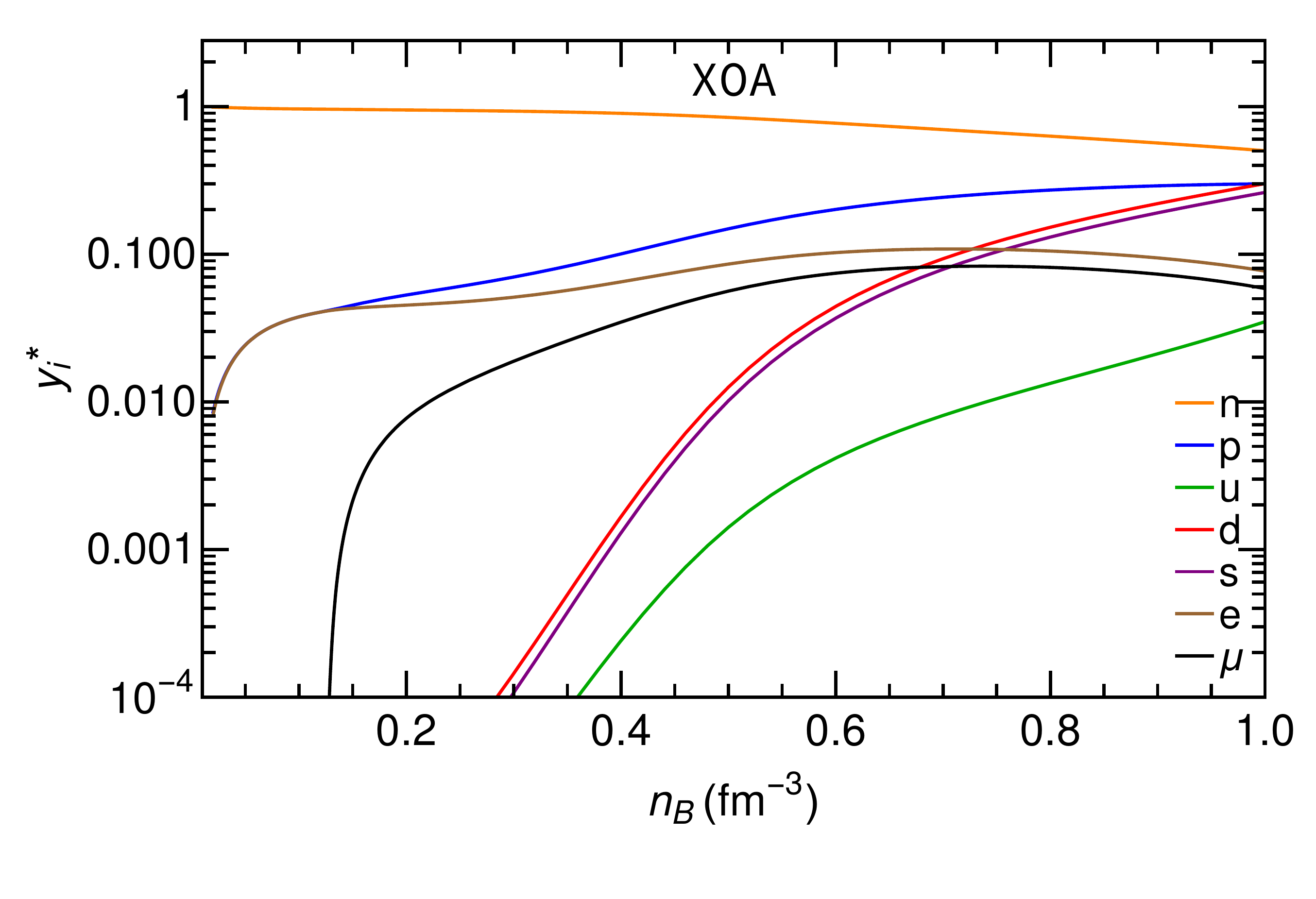}
\caption{Particle fractions of the crossover model XOA versus baryon density. Quarks are still present below 0.3 fm$^{-3}$ but at two orders of magnitude less than what is shown here.} 
\label{fig:YIA}
\end{figure}

\begin{figure}[h!]
\includegraphics[scale=0.32]{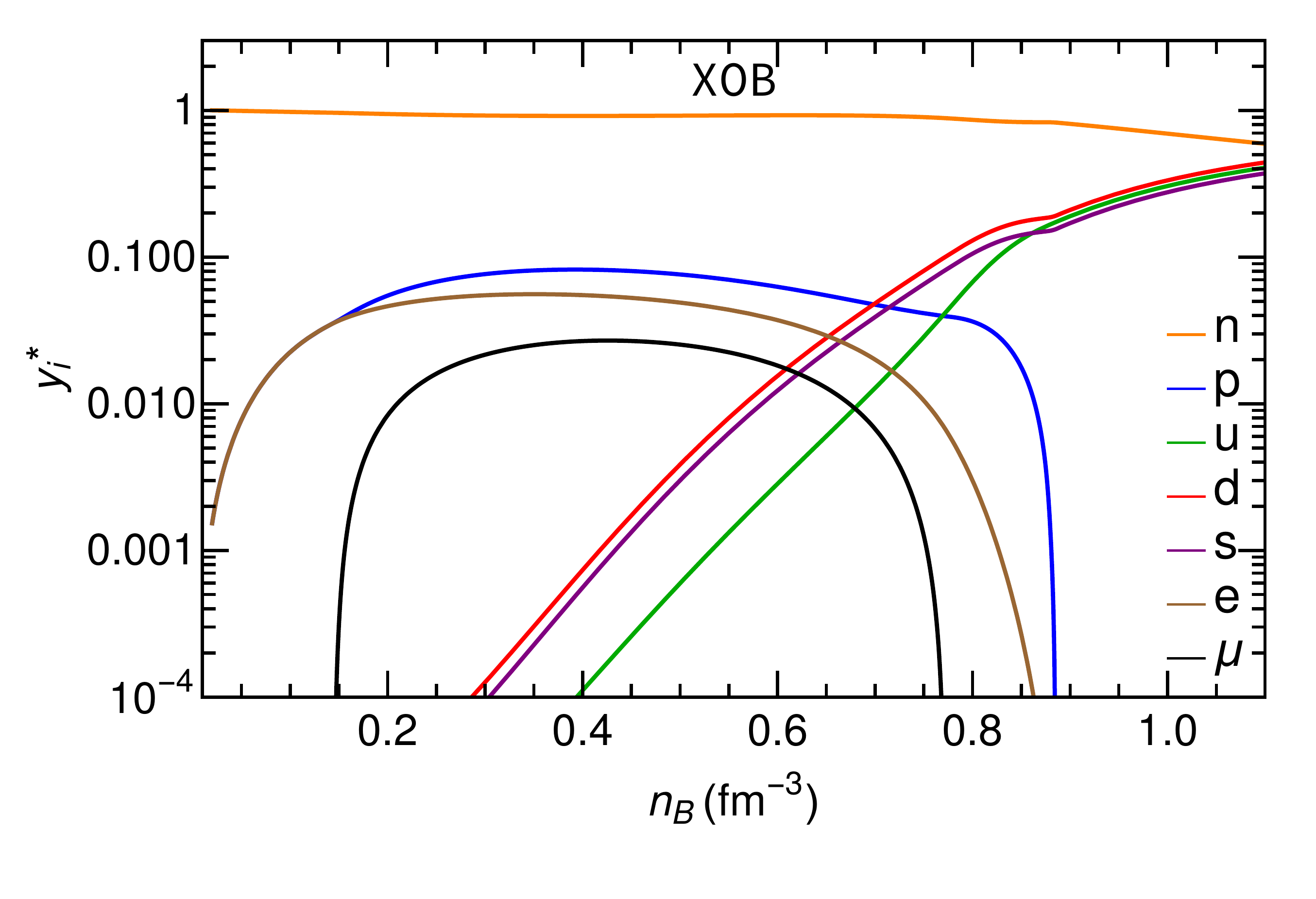}
\caption{Particle fractions of model XOB versus baryon density. Muons drop out of the system at 0.77 fm$^{-3}$ and protons at 0.88 fm$^{-3}$ leading to a sharp peak in the adiabatic sound speed (cf. Fig.~\ref{fig:OOC}). Electrons are still present above 0.85 fm$^{-3}$ but at two orders of magnitude less than what is shown here.} 
\label{fig:YIB}
\end{figure}

\begin{figure}[h!]
\includegraphics[scale=0.32]{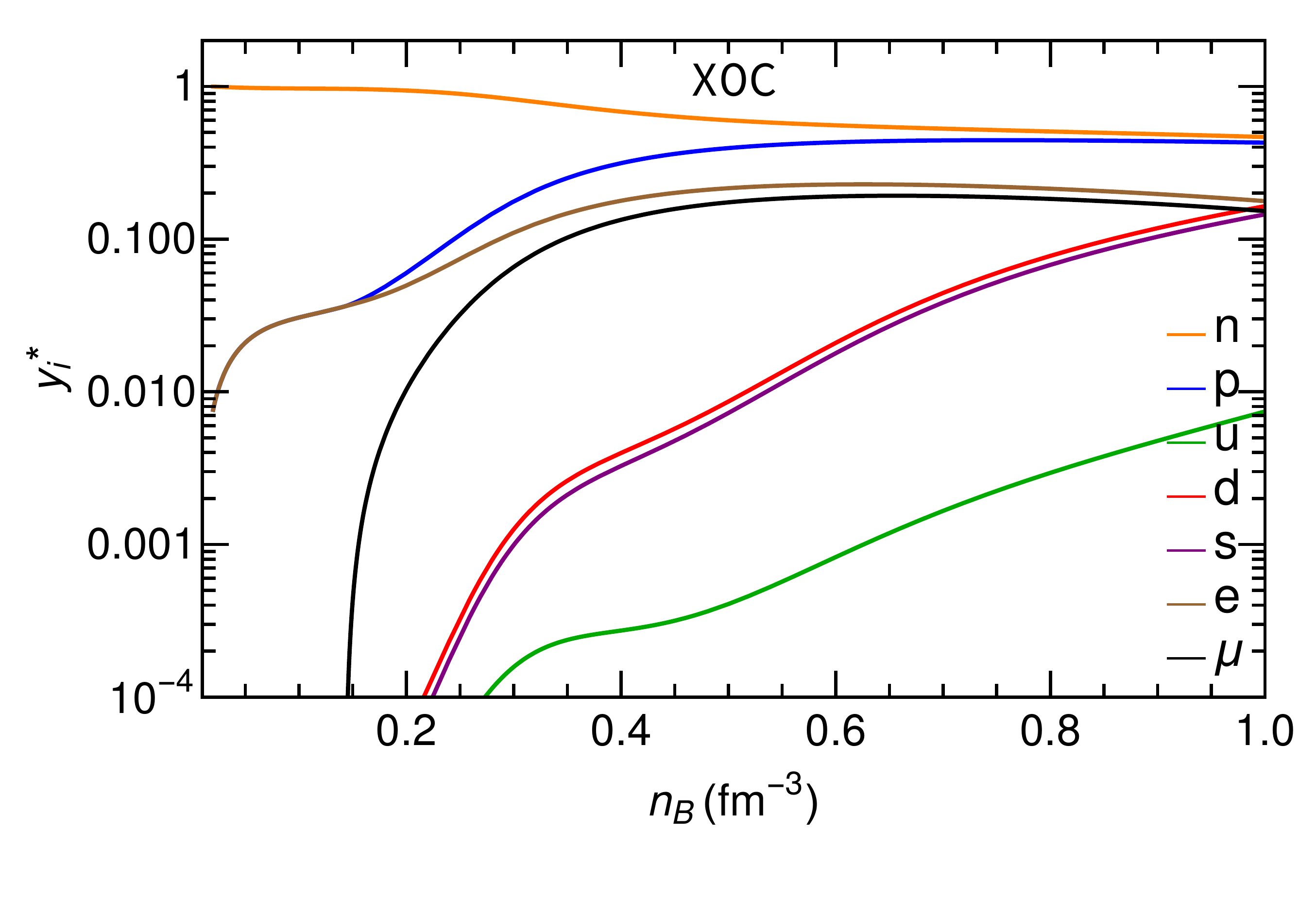}
\caption{Particle fractions of model XOC versus baryon density. The inflectionary behavior of the quark and neutron fractions around 0.4 fm$^{-3}$ is responsible for the broad peak behavior of XOC occurring in Fig.~\ref{fig:OOC}.}
\label{fig:YIC}
\end{figure}

Figure~\ref{fig:OOC} shows the difference of the inverses of the adiabatic and equilibrium sound speeds, $\Delta(c^{-2})\equiv 1/\csqeq-1/\csqad$ as a function of the baryon density. This quantity is particularly important in the context of \gms~because it enters directly in the calculation of the \bv~frequency (discussed in more detail in the next section). 
\vspace{3mm}

A comparison between this figure and Figs.~\ref{fig:YIA}--\ref{fig:YIC} which show the particle fractions corresponding to the three crossover models used in this work, reveals a direct correlation between sharp maxima in the former and particle appearance/disappearance in the latter. 
Smooth maxima, such as those exhibited by XOA around 0.7 fm$^{-3}$ and XOC around 0.4 fm$^{-3}$, reflect non-monotonic behaviors in the slopes of the quark and the neutron fractions. Also worth noting is that, in the present framework and with the chosen parametrizations, quarks are never the dominant contributors to the baryon density for densities relevant to neutron stars. 
\vspace{3mm}

\begin{table}[h]
\caption{Various representations of the quark content of $2\,\Msolar$ NSs corresponding to the Gibbs and the three crossover models in this paper. The symbols are 
$M_{\rm B}$: the total baryonic mass, 
$Y_Q^{\rm{bar}}$: contribution of quarks to the total baryon number/total baryon number, 
$Y_Q^{\rm{part}}$: quark particle number/total particle number,
$Y_Q^{\rm{nuc}}$: quark particle number/nucleon number, 
${M_Q^{\rm{B}}}/{M_{\rm B}}$: contribution of quarks to the total baryonic mass/total baryonic mass, and 
${M_Q^{\rm{G}}}/{M_{\rm G}}$: contribution of quarks to the total gravitational mass/total gravitational mass. 
} 
\begin{center} 
\begin{tabular}{ccccccc}
\hline
\hline
Model &  $M_{\rm B}$ & $Y_Q^{\rm{bar}}$  & $Y_Q^{\rm{part}}$  & $Y_Q^{\rm{nuc}}$   & ${M_Q^{\rm{B}}}/{M_{\rm B}}$  & ${M_Q^{\rm{G}}}/{M_{\rm G}}$  \\ 
& $(\Msolar)$  & ($\times 10^{-2}$) & ($\times 10^{-2}$) & 
($\times 10^{-2}$) & ($\times 10^{-2}$) & ($\times 10^{-2}$) \\ \hline
XOA    & 2.31  &  0.35  &  1.03  &  1.04  &  0.35 &   0.50  \\
XOB    & 2.33  &  2.15  &  6.18  &  6.58  &  2.15 &   3.04 \\ 
XOC    & 2.29  &  0.06  &  0.19  &  0.19  &  0.06 &   0.09 \\
Gibbs  & 2.35  &  1.91  &  5.53  &  5.85  &  1.91 &   2.89 \\
\hline \hline
\end{tabular}
\end{center}
\label{tab:quarkfrac}
\end{table}

Consequently, the contributions of quarks to the total baryon number as well as the total mass of the star (both baryonic and gravitational) are rather small in the models considered, as shown in Table~\ref{tab:quarkfrac} for stars with gravitational mass $M=2\,\Msolar$. These are straightforwardly calculated as follows: The solution of the TOV equations gives, among other things, the baryon density as a function of the star's radius, $\nb(r)$. The quark particle fractions as functions of the baryon density are obtained from the $\beta$-equilibrated equivalents of Eqs. (\ref{qno}) and (\ref{qnoG}) for the KW and Gibbs cases, respectively. Explicitly, $y_q (r) = y_{q,\beta}^*[\nb(r)]$. For the quark baryon fraction at each radius we divide this by 3. The quark particle densities are then $n_q(r) = y_q(r)\nb(r)$. The baryon number due to quarks of all species $q=u,d,s$ in the star is given by the integral 
\begin{equation}
N_Q = 4\pi \int_0^R dr r^2 \frac{\sum_q n_q(r)/3}{[1-2GM(r)/r]^{1/2}}.
\end{equation} 
Here, $M(r)$ is the total gravitational mass of the star as a function of its radius $r$, also given by the solution of the TOV equations. Therefore, the amount of baryonic mass in the star provided by quarks is $M_Q^{\rm{B}} = m_H N_Q$~\cite{glendenning2012compact}. For the quark gravitational mass we begin by calculating the quark energy density as a function of the radius according to $\ep_Q(r) = \ep_{Q,\beta}^*[\nb(r)]$ where $\ep_Q^* = -S P_Q + \sum_q n_q^*\mu_q$. Afterwards, we perform the integral $M_Q^{\rm{G}} = 4\pi \int_0^R dr r^2 \ep_Q(r)$. 
\vspace{3mm}

Specifically, in model XOA, quarks contribute less than 0.5\% of the total baryon number, about 1\% of particles in its $2\,\Msolar$ stars are quarks and they are responsible for around 0.35\% (0.5\%) of the total baryonic (gravitational) mass of the star. For XOB and Gibbs models, quarks contribute about $2\%$ of the total baryon number, over 5\% of the total particle number, and 2\% (3\%) of the total baryonic (gravitational) mass of the respective $2\,\Msolar$ stars. In model XOC, quarks are, for all intents and purposes, irrelevant. 
As noted before, such a clean separation may not be possible in treatments that intermingle hadron/nucleon and quark interactions.

\subsection{Sound speeds and the \bv~frequency} 

Having outlined the thermodynamics of the multi-component system and the specific models employed in our study of the smooth crossover transition in neutron stars, we turn now to the calculation of the star's \gm~frequencies.
Our main goal is to compare the behavior of the \gm~frequencies in the crossover model with those in the Gibbs mixed phase.
The \gm~frequencies ($\nu_g$=$\omega/(2\pi)$) and normalized amplitudes for the radial and tangential parts of the fluid perturbation ($\xi_r$ and $\xi_h$ respectively) are estimated within the relativistic Cowling approximation (see below) by computing numerical solutions to the following equations of motion for fluid variables $U,V$~\cite{Jaikumar:2021jbw}
\beq
\label{eq:uv}
\frac{dU}{dr}&=&\frac{g}{\csqad}U+{\rm e}^{\lambda/2}\left[\frac{l(l+1){\rm e}^{\nu}}{\omega^2}-\frac{r^2}{\csqad}\right]V \nn \\ 
\frac{dV}{dr}&=&{\rm e}^{\lambda/2-\nu}\frac{\omega^2-N^2}{r^2}U+g\Delta(c^{-2})V \,,
\eeq
which are simplified forms of the original perturbation equations~\cite{McD83,RG92,Kantor:2014lja}. 
In \Eqn{eq:uv}, $U$ = $r^2{\rm e}^{\lambda/2}\,\xi_r$, $V$ = $\omega^2 r\, \xi_h$ = $\delta P/(\ep+P)$,   $\Delta(c^{-2})=\ceq^{-2}-\cad^{-2}$, $\lambda$ and $\nu$ are metric functions. The scale of the mode frequency is set by the \bv~frequency 
\beq
N^2 = g^{2}\Delta(c^{-2}){\rm e}^{\nu-\lambda},
\eeq
where $g=-\nabla P/(\ep+P)$. 

The relativistic Cowling approximation neglects the back reaction of the gravitational potential by excluding metric perturbations that must accompany matter perturbations in a general relativistic treatment~\cite{Thorne:1967a,Thorne:1967b,Lindblom:1983,Detweiler:1985,Finn:1987,Andersson:1995wu}. 
It reduces the number and complexity of the equations we have to solve, while providing results for \gm~frequencies that are accurate at the few \% level~\cite{gregorian2015nonradial}. 
Details on the solution methods for~\Eqn{eq:uv} and relevant boundary conditions are provided in Ref.~\cite{Jaikumar:2021jbw}. 
\vspace{3mm}

\begin{figure}[h!]
\includegraphics[scale=0.32]{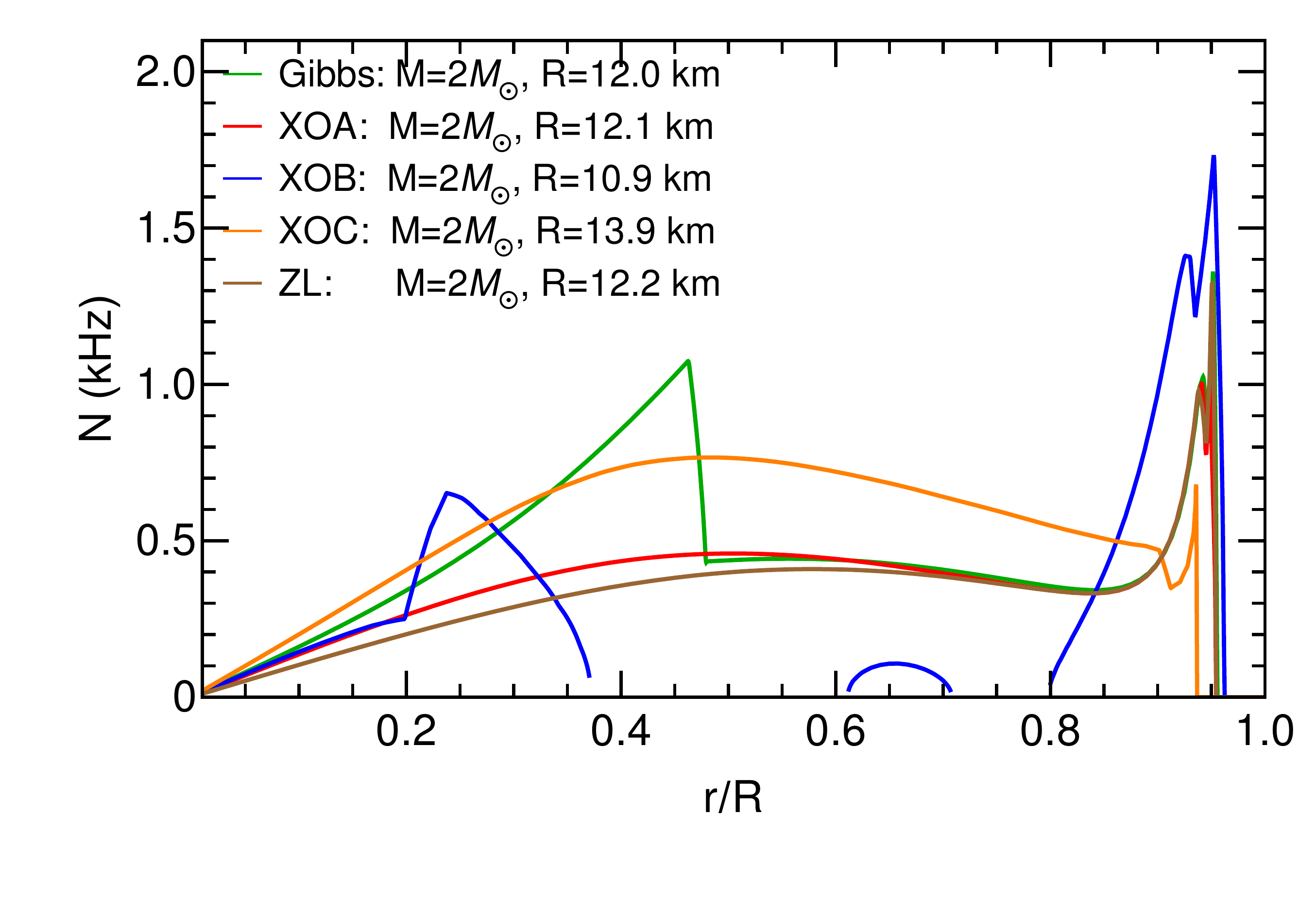}\\[-3.0ex]
\caption{The \bv~frequency in a hybrid star of mass $2.0\,\Msolar$ for the Gibbs and crossover models. 
The \bv~frequency for the ``good'' crossover model XOA is very similar to the nucleonic ZL EOS (which includes muons), whereas the Gibbs model shows a distinct peak corresponding to the rapid onset of quark matter. Parameters for the nuclear and quark EOSs are as in Table~\ref{tab:Parameters}. 
}
\label{fig:BV-EOS}
\end{figure}

\begin{figure}[h!]
\includegraphics[scale=0.32]{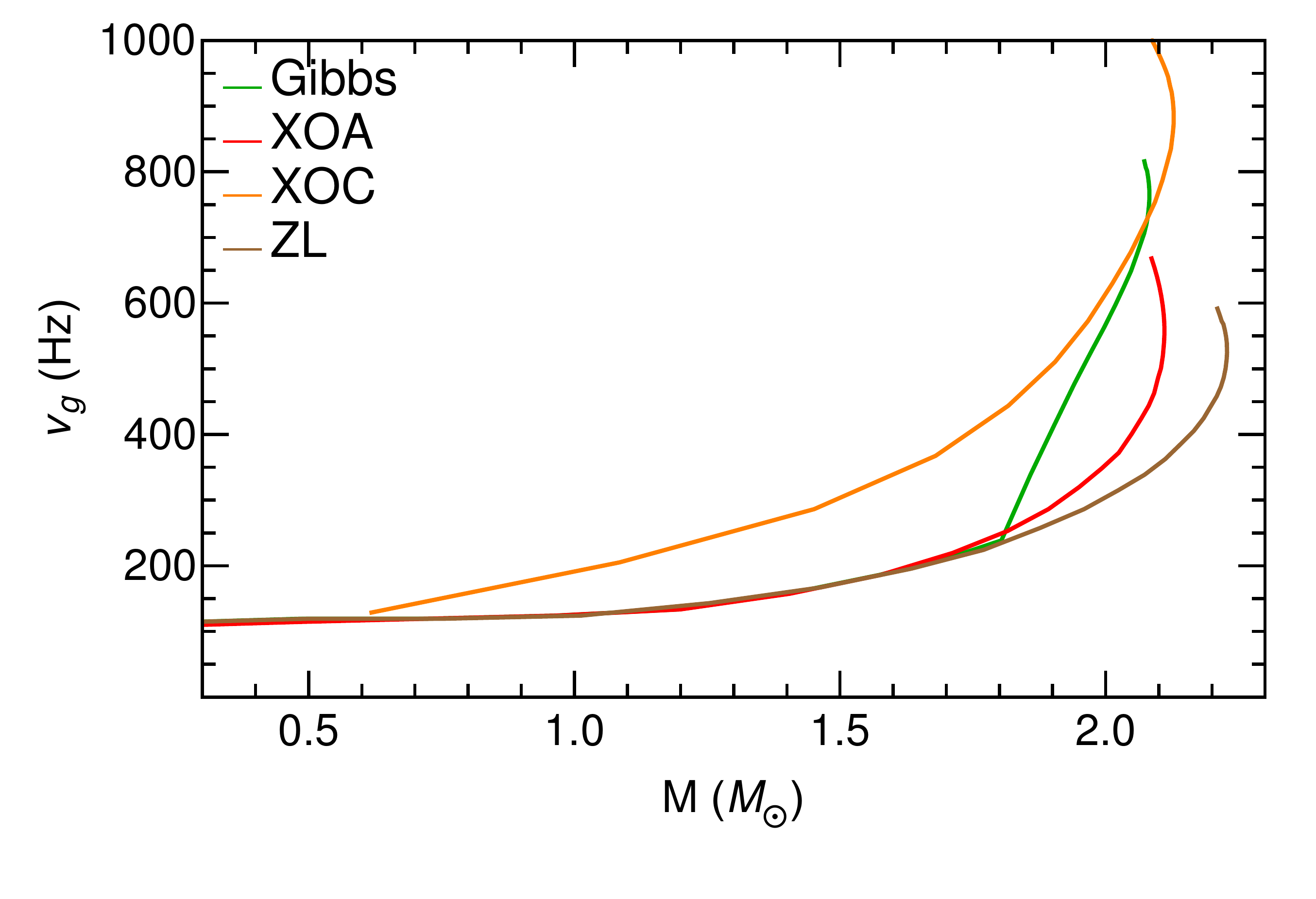}\\[-3.0ex]
\caption{The \gm~frequency as a function of the stellar mass in the Gibbs, crossover, and ZL models. 
Parameters for the nuclear and quark EOSs are as in Table.~\ref{tab:Parameters}. 
The g-modes corresponding to XOB are unstable and therefore, this model, is excluded from the present and the next two figures.} 
\label{fig:gmode-EOS}
\end{figure}

\begin{figure}[]
\includegraphics[scale=0.28]{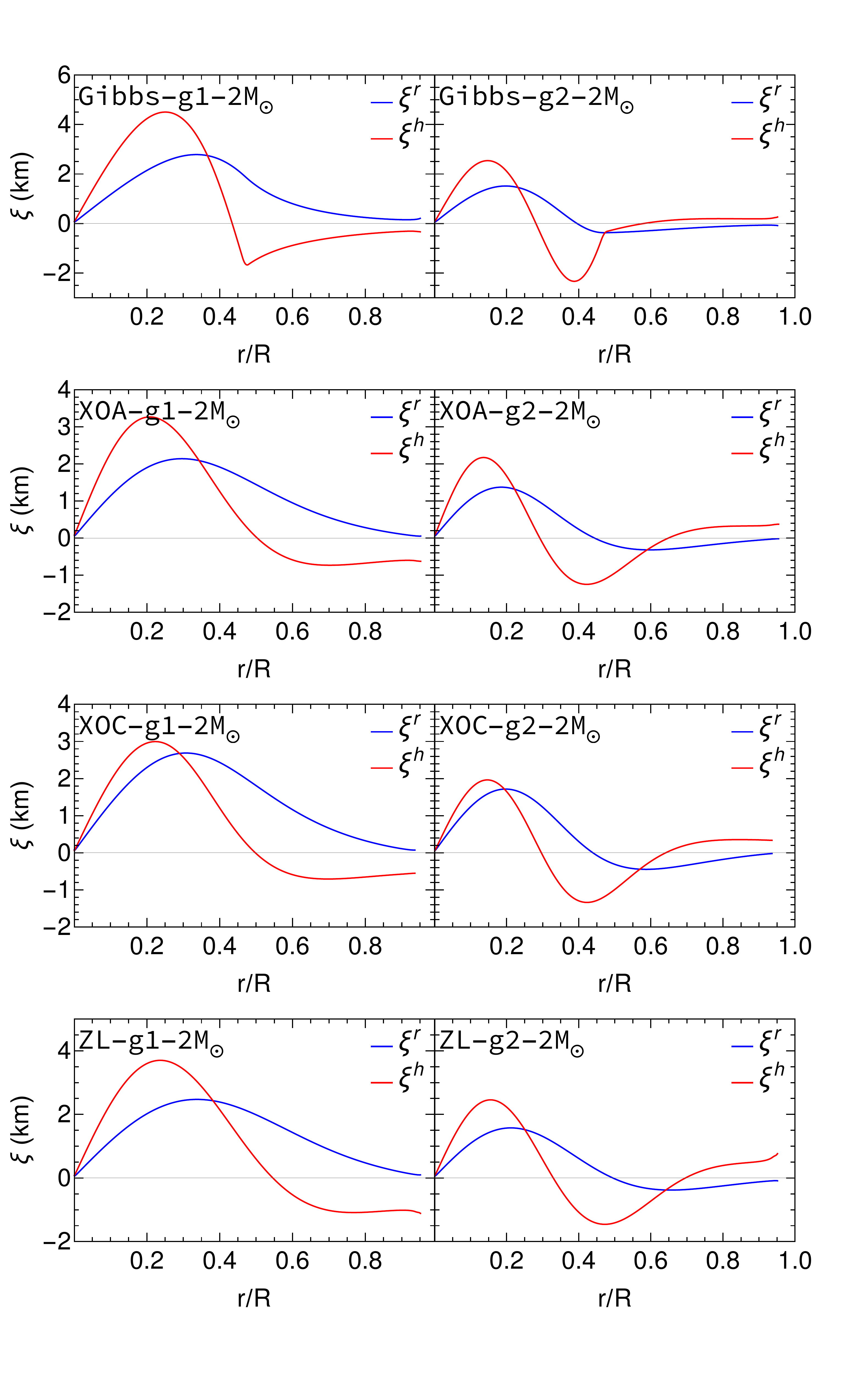}\\[-3.0ex]
\caption{Amplitudes of the radial ($r$) and transverse ($h$) components of the \gm~displacement (eigenfunctions) as a function of distance from the center for the ZL, XOA, XOC and Gibbs EOSs for a $\sim2\,\Msolar$ star. The order of the mode (g1 is fundamental, g2 is overtone) is indicated in the legend in the panels. 
}
\label{fig:eigenfunctions}
\end{figure}

\begin{figure}[h!]
\includegraphics[scale=0.32]{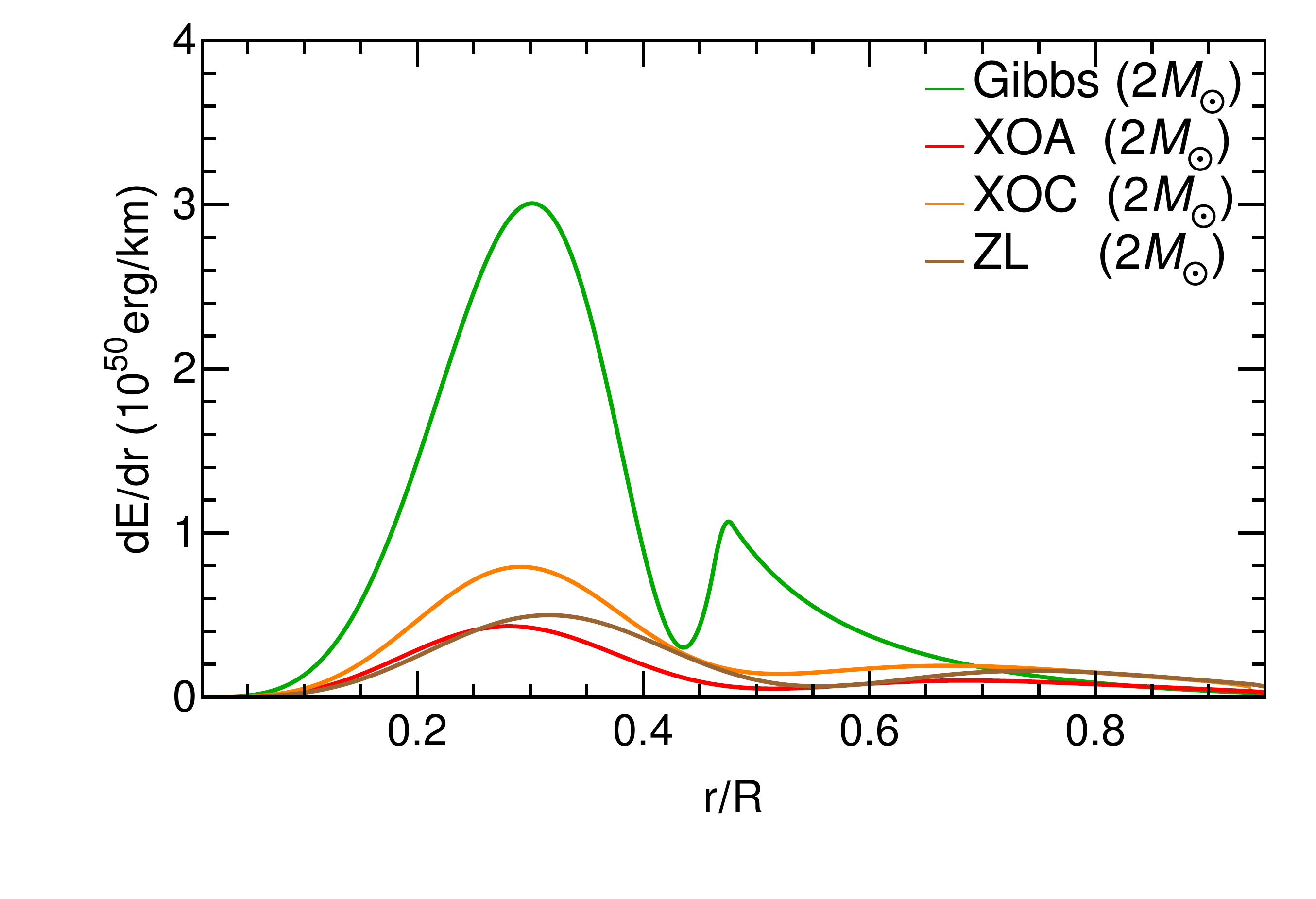}\\[-3.0ex]
\caption{Energy/unit distance of the fundamental \gm~as a function of distance from the center for the ZL, XOA, XOC, and Gibbs EOS for a  $\sim2\,\Msolar$ star. Note that the mode energies in the ZL and crossover cases are scaled up by a factor of 10 for the purpose of comparison.}
\label{fig:energy}
\end{figure}

Figure~\ref{fig:BV-EOS} is a comparison of the \bv~frequency in the three models considered in this work. The crossover model, where quarks are always present in the core EOS, albeit in minuscule fractions, resembles the purely nucleonic ZL EOS in this respect, while the sudden onset of quarks in the Gibbs model is clearly imprinted in the form of a sharp peak. 
Quarks enter at a density $\nb \simeq$ 0.514 fm$^{-3}$ corresponding to $r/R$ = 0.473 ($r=0$ at the center) in the Gibbs model. As a consequence of the difference of sound speeds being negative in distinct density regimes for XOB (Fig.~\ref{fig:OOC}), the corresponding \bv~frequency is imaginary, implying an instability to convection.\footnote{This feature could be an artifact of the Cowling approximation, but will likely be absent in the solution of the full general relativistic treatment~\cite{Thorne:1967a,Thorne:1967b,Lindblom:1983,Detweiler:1985,Finn:1987,Andersson:1995wu} 
of the \gm~in which $\csqeq$ does not enter explicitly.} 
However, convection is absent at zero temperature; therefore these regions are unphysical and can play no role in the global \gm~spectrum. Accordingly, XOB is omitted from Figs.~\ref{fig:gmode-EOS}, \ref{fig:eigenfunctions} and \ref{fig:energy}.

Figure~\ref{fig:gmode-EOS} compares the \gm~frequency, $\nu_g=\omega/(2\pi)$, for the crossover and Gibbs models. While XOA and XOC are very similar to the nucleonic ZL EOS (which includes muons), the Gibbs model shows a distinctly rising spectrum corresponding to the rapid onset of quark matter. These findings are consistent with the result for the \bv~frequency in Fig.~\ref{fig:BV-EOS} and the conclusions in Paper I. 
In the crossover model XOB, protons disappear above some critical density. In contrast to the smooth behavior of \gm~frequencies in XOA and XOC, the sudden disappearance of protons in XOB produces a sharp rise in the spectrum akin to the Gibbs case but renders \gms~to become unstable (not shown in Fig.~\ref{fig:gmode-EOS}), confirming that dramatic changes in the \gm~frequency require {\it appearance or disappearance of a (strongly interacting) particle species, not merely a smooth change in composition}. 
This is why, except for extreme parameter choices, crossover models will not show the \gm~feature resulting from the presence of quarks that Gibbs models do.

The panels in Fig.~\ref{fig:eigenfunctions} show the comparison of the core \gm~amplitude between the three chosen models (ZL, crossover and Gibbs). The radial component $\xi_r$ of the fundamental mode (labeled ``g1'' in the panels) has no nodes in the core, while the radial part of the first overtone (labeled ``g2'') has one, as expected. The horizontal component $\xi_h$ has one more node than the corresponding radial component of the same order.
The larger amplitude of $\xi_h$ relative to $\xi_r$ indicates that the \gm~is dominated by transverse motion of the perturbed fluid. While there is little difference between the ZL and crossover models in the profile of these eigenfunctions, the Gibbs case is markedly different. Its amplitude relative to the other two is larger, and it changes abruptly upon the onset of quark matter in the core.

The energy per unit radial distance $dE_T/dr$ contained in the oscillatory motion corresponding to a frequency $\omega$ is given in terms of the amplitude as~\cite{McD83} 
\be
\frac{dE_T}{dr} =\frac{\omega^2\,r^2}{2}(\ep+P){\rm e}^{(\lambda-\nu)/2}\,\left[\xi_r^2{\rm e}^{\lambda}+l(l+1)\xi_h^2\right] \,.
\ee

Figure~\ref{fig:energy} shows the comparison of the core \gm~energy/unit distance for the three chosen models (ZL, crossover and Gibbs) for a $\sim2\,\Msolar$ star. The typical scale of the energy/unit distance deep in the core is approximately $10^{50}\, \rm{ergs/km}$ for the ZL and crossover models, while it is of order $10^{51}\, \rm{ergs/km}$ for the Gibbs case. 
While the profiles are similar for the first two, the mode energy in the Gibbs case is overwhelmingly larger in the core, once quark matter appears. Thus, both the frequency and the amplitude (and hence the energy) of the core \gm~is strongly amplified in quark matter in comparison to nucleonic matter or weaker forms of the phase transition. This can have bearing on the gravitational wave detection of \gms~excited in neutron star mergers as discussed below.

\subsection{Discussion}
\label{sec.Discs}

The results in the previous subsection raise some points that are noteworthy. In particular, they highlight the relation between the behaviors of the speeds of sound and of the particle concentrations, and the spectrum of the \gm~signal. Thus, the latter becomes a diagnostic which distinguishes nucleonic and hybrid matter \textit{as well as} the Gibbs and crossover transitions.

Peaks in the sound-speed difference $\delta c \equiv \csqad - \csqeq$, and the difference of the inverses $\Delta (c^{-2}) \equiv 1/\csqeq - 1/\csqad$ occur when particles appear or disappear [i.e. when the number of degrees of freedom (DOF) of the system changes] and when the particle concentrations $y_i(\nb)$ are not monotonic or have inflection points (i.e., their first or second derivatives with respect to $\nb$ change sign).

The introduction or removal of a particle species from the system occurs when the relevant chemical potential either exceeds or falls below its rest mass threshold while maintaining charge neutrality. 
Whereas the functional form of the $y_i$'s depends on the parametrization of the EOS, a change in the number of DOFs also leads to non-monotonic or inflectional behavior to the concentrations of particles already present in the system. Thus, in some sense, specific choices of the EOS parameters can mimic aspects of the emergence of new particles that are relevant to the sound speeds.

Signals of \gm~with a characteristic fast rise in its frequency (such as 
those corresponding to Gibbs shown in Fig.~\ref{fig:gmode-EOS} and XOB for which the \gm~becomes unstable) 
can occur only when particle species enter or leave the system. In such cases, the peaks in $\Delta (c^{-2})$ are sharp and asymmetric: vertical rise and quasi-lorentzian decay for appearance (see Fig.~\ref{fig:OOC}, Gibbs), and quasi-lorentzian rise and vertical drop for disappearance (XOB); whereas peaks due to parametrizations resemble symmetric Gaussians (XOC). Therefore, this kind of signal cannot be produced by quarks in  matter with a smooth crossover because quarks are always present: their concentrations are vanishingly small at lower densities but never identically zero. 

It is, however, possible to parametrize the hadronic EOS such that, in $\beta$-equilibrated crossover matter, protons exit the system. The magnitude of the peak of $\Delta (c^{-2})$ \textit{appears} to be proportional to the number of remaining DOFs (smallest in PNM; largest in $n$-$uds$-$e$ matter). This produces \gm~frequency spectra similar to those found in matter with a Gibbs construction. So, although quarks are not directly responsible for this effect, they do serve the purpose of amplifying it.

For the specific case of crossover model XOB, a rather extreme parametrization ($K=260$ MeV, $S_v=30$ MeV, $L=70$ MeV, $\gamma_1=1.1$) was required for the proton disappearance while also meeting neutron-star constraints and having a peak at low-enough densities to be relevant. Consequently, the tentative conclusion is that hyperons or a first-order transition into quark matter through Gibbs construction are more likely to cause a distinctive peak in the \bv~frequency than a crossover transition; nevertheless, the latter remains a viable, if improbable, option. 

It is pertinent to mention that we have not performed calculations for first-order transitions with a Maxwell construction, which assume a sufficiently large surface tension between the pure hadronic and quark phases in bulk separated by a sharp boundary. 
It has been shown that in such cases, \gms~always vanish when the perturbed fluid element adjusts instantaneously (i.e. a very rapid conversion) to maintain thermodynamic equilibrium, but can arise if the microscopic phase conversion at the interface is slow enough, leading to distinctive features on the extended hybrid branch of NSs~\cite{Pereira:2017rmp,Tonetto:2020bie}. Such \gms~associated with a discontinuity in density (``\textit{discontinuity} \gms'', different from the ``\textit{compositional} \gms'' that we consider) 
has been widely studied in the literature; see e.g., extensive discussions in Refs.~\cite{Sotani:2001bb,Flores:2013yqa,Ranea-Sandoval:2018bgu,Rodriguez:2020fhf,Lau:2020bfq}.

\section{Summary and Conclusions}
\label{sec:Concs}

The main objectives of this work were to examine \gm~frequencies in a smooth crossover scenario of the hadron-to-quark transition, and to compare them with those of a first-order transition treated using Gibbs and Maxwell constructions in Paper I~\cite{Jaikumar:2021jbw}. 
For the crossover model, we chose the recent approach adopted by Kapusta and Welle~\cite{Kapusta:2021ney}, who constructed an EOS for PNM that resembled the smooth crossover observed in lattice calculations at finite temperatures. 
We have generalized their approach to $\beta$-equilibrated NSM so that comparisons with the results of Paper I could be made. To describe nucleons, we used the ZL parametrization~\cite{Zhao:2020dvu} which reproduces near-saturation laboratory data as well as results of chiral EFT calculations~\cite{Drischler:2020hwi,Drischler:2020fvz} up to $\sim$ $2.0\,\nsat$. 
For quark matter we used the vMIT model~\cite{Gomes:2018eiv} with repulsive interactions. Results of our crossover EOSs tally with observational findings of the radii of $\sim\,1.4$ and $\sim\,2.0\,\Msolar$ NSs~\cite{Miller:2019cac,Riley:2019yda,Miller:2021qha,Riley:2021pdl}.  
Calculations of the equilibrium and adiabatic speeds of sound were performed following the procedures developed in Paper I. Our work here is focused on zero temperature and does not consider superfluidity in either nucleons or quarks. Inclusion of these effects will be taken up in a future work.

The results of the amplitudes of the \gm~frequencies and their associated amplitudes of the gravitational energy radiated for the chosen hadron-to-quark crossover models lie between those of the first-order phase transitions that employ Maxwell and Gibbs constructions. 
For the case of the Maxwell construction, the transition region is devoid of \gm~frequencies as the equilibrium and adiabatic sound speeds both vanish. Consequently, \gm~oscillations are permitted only for the pure nucleonic and quark phases. Their amplitudes in these regions are, however, rather small. 
In contrast, the mixed phase in the case of the Gibbs construction yields amplitudes of \gm~frequencies and the energy radiated that are significantly larger than those of the crossover model. 

We note that \gm~frequencies can be exceptionally large in the presence of superfluidity ($\approx$ 750 Hz for a hyperonic star~\cite{Dommes:2015wul} and $\approx$ 450 Hz for a nucleonic star~\cite{Gusakov:2013eoa,Passamonti:2015oia}), similar to the results for the Gibbs mixed phase. However, the reason for the enhancement in the two cases is different. In the case of superfluidity, the temperature-dependence of the entrainment terms serves to increase the \gm~frequency at typical neutron star temperatures, while in the Gibbs case, the enhancement is purely composition-dependent. 
A study of resonant excitations of such superfluid modes in coalescing neutron star binaries~\cite{Yu:2016ltf} suggests that the amplitude of these modes is weaker by a factor of 20 or so, compared to modes from the normal fluid. Note that the \gms~we discuss here are also higher in frequency compared to the low-frequency \gms~($\sim 50$ Hz) that might strain the neutron star crust to breaking point and lead to precursor flares in gamma-ray bursts~\cite{Kuan:2021jmk}.

The relatively larger frequency, amplitude and energy of the \gm~in the Gibbs case inferred from 
Figs.~\ref{fig:gmode-EOS}-\ref{fig:energy} 
have observational implications for gravitational waves from neutron star mergers. It has been established that \gms~can couple to tidal forces and draw energy and angular momentum from the binary to the neutron star, leading to an accelerated merger and a concomitant phase shift in the gravitational waveform~\cite{Lai:1993di}. This coupling will be largest for the Gibbs case, with its higher energy at resonance, and also because higher resonance frequencies are excited later in the inspiral, 
when tidal forces are strongest. Estimates of the resulting phase shift were presented in Paper I (Eq. (89)) and found to be comparable to that from \gms~in ordinary neutron stars (due to longer merger times) within uncertainties arising from the value of the tidal coupling. As these uncertainties are reduced through improved theoretical calculations, the case of a hybrid star may be distinguished from an ordinary neutron star. 
We also infer from our results that were high-frequency \gms~to be detected in upgraded LIGO and Virgo observatories, it would indicate a first-order phase transition akin to a Gibbs construction. In light of data from GW170817, lower bounds on the excitation of non-radial oscillations in binary mergers~\cite{Pratten:2019sed} affirm that a third generation network with its improved sensitivity and larger bandwidth can shed new light on the composition of the neutron star core. 
\vspace{4mm}

\begin{acknowledgments}

C.C. acknowledges support from the European Union's Horizon 2020 research and innovation programme under the Marie Sk\l{}odowska-Curie grant agreement No. 754496 (H2020-MSCA-COFUND-2016 FELLINI).
S.H. is supported by the National Science Foundation, Grant PHY-1630782, and the Heising-Simons Foundation, Grant 2017-228. 
P.J. is supported by the U.S. National Science Foundation Grant No. PHY-1913693.
M.P.'s research was supported by the Department of Energy, Grant No. DE-FG02-93ER40756.

\end{acknowledgments}

\bibliography{gmode_paper}
\end{document}